\documentclass[12pt]{article}
\title{Multi-black holes from nilpotent Lie algebra orbits}
\usepackage{mathrsfs}
\usepackage{amsmath}

\global\arraycolsep=1pt
\usepackage[colorlinks=true,backref=true,linkcolor=black,anchorcolor=black,citecolor=black,filecolor   =black,menucolor=black,pagecolor=black,urlcolor=black]{hyperref}
\usepackage[Symbol]{upgreek}
\usepackage{bbm}
\usepackage{dsfont}
\usepackage{amssymb}
\usepackage{textcomp}
\usepackage{wasysym}

\newcommand{\Scal}[1]{\Bigl ({#1} \Bigr )}
\newcommand{\scal}[1]{\bigl ({#1} \bigr )}
\def\bea{\begin{eqnarray}}
\def\eea{\end{eqnarray}}
\def\be{\begin{equation}}
\def\ee{\end{equation}}
\newcommand{\CR}{\nonumber \\*}

\newcommand{\trace}{\hbox {Tr}~}

\DeclareMathAlphabet{\mathpzc}{OT1}{pzc}{m}{it}

\DeclareMathOperator{\ad}{ad}

\newcommand{\ord}[1]{{\scriptscriptstyle (#1)}}

\def\C{\mathscr{C}}

\usepackage{graphicx}

\def\un{{\mathpzc{1}}}
\def\deux{{\mathpzc{2}}}
\def\trois{{\mathpzc{3}}}

\def\cinq{{\mathpzc{5}}}

\def\DJo{$\;$\kern-.4em \hbox{D\kern-.8em\raise.15ex\hbox{--}\kern.35em okovi\'c}}

\def\ie{{\it i.e.}\ }

\def\nn{\nonumber}

\def\N{\mathcal{N}}

\def\ft#1#2{\tfrac{#1}{#2}}

\def\C{{\mathscr{C}}}
\def\V{{\mathcal{V}}}
\def\w{{\scriptstyle W}}

\def\G{{\mathfrak{G}}}
\def\H{{\mathfrak{H}}}
\def\P{{\mathfrak{P}}}
\def\M{{\mathcal{M}}}

\def\R{{\mathcal{R}}}

\def\m{{\mathpzc{m}}}
\def\n{{\mathpzc{n}}}
\def\p{{\mathpzc{p}}}
\def\s{{\mathpzc{s}}}

\def\zero{{\mathpzc{0}}}
\def\un{{\mathpzc{1}}}
\def\deux{{\mathpzc{2}}}
\def\trois{{\mathpzc{3}}}

\def\invo{{\APLstar}}

\def\DJo{$\;$\kern-.4em \hbox{D\kern-.8em\raise.15ex\hbox{--}\kern.35em okovi\'c}}
\def\Ic{{I\hspace{-0.6mm}c}}

\newcommand{\eprint}[1]{{\href{http://arxiv.org/abs/#1}{\texttt{[#1}]}}}
\newcommand{\eprintN}[1]{{\href{http://arxiv.org/abs/#1}{\texttt{#1 [hep-th]}}}}

\def\e{\boldsymbol{e}}

\def\h{\boldsymbol{h}}

\def\g{\mathfrak{g}}
\def\G{\mathfrak{G}}
\def\h{\mathfrak{h}}
\def\H{\mathfrak{H}}

\def\ft#1#2{${\textstyle{{\scriptstyle #1}\over {\scriptstyle #2}}}$}

\def\w{{\scriptstyle W}}

\def\gl{\mathfrak{gl}}
\def\sl{\mathfrak{sl}}
\def\so{\mathfrak{so}}
\def\su{\mathfrak{su}}

\def\e{\mathfrak{e}}

\def\Ha{\mathcal{H}}

\def\nn{\nonumber}

\def\N{\mathcal{N}}

\begin{document}
\allowdisplaybreaks[1]
\renewcommand{\thefootnote}{\fnsymbol{footnote}}
\numberwithin{equation}{section}
\def\corr{$\spadesuit \, $}
\def\trefle{$ \, $}
\def\kscorr{$\diamondsuit \, $}
\begin{titlepage}
\begin{flushright}
\
\vskip -2.5cm
{\small AEI-2009-056}\\
\vskip 1cm
\end{flushright}
\begin{center}
{\Large \bf
Multi-black holes from nilpotent Lie algebra orbits}
\\
\lineskip .75em
\vskip 3em
\normalsize
{\large  Guillaume Bossard\footnote{email address: bossard@aei.mpg.de} and
Hermann Nicolai\footnote{email address: Hermann.Nicolai@aei.mpg.de}}\\
\vskip 1 em
$^{\ast\dagger}${\it AEI, Max-Planck-Institut f\"{u}r Gravitationsphysik\\
Am M\"{u}hlenberg 1, D-14476 Potsdam, Germany}
\\

\vskip 1 em
\end{center}
\begin{abstract}
{\footnotesize
For $\N\ge 2$ supergravities, BPS black hole solutions preserving four
supersymmetries can be superposed linearly, leading to well defined 
solutions containing an arbitrary number of such BPS black holes at 
arbitrary positions. Being stationary, these solutions can be understood 
via associated non-linear sigma models over pseudo-Riemaniann spaces 
coupled to Euclidean gravity in three spatial dimensions. As the main
result of this paper, we show that whenever this pseudo-Riemanniann space 
is an irreducible symmetric space $\G/\H^*$, the most general solutions 
of this type can be entirely characterised and derived from the nilpotent 
orbits of the associated Lie algebra $\g$. This technique also permits the 
explicit computation of non-supersymmetric extremal solutions which 
cannot be obtained by truncation to $\N=2$ supergravity theories. 
For maximal supergravity, we not only recover the known BPS solutions 
depending on $32$ independent harmonic functions, but in addition find 
a set of non-BPS solutions depending on $29$ harmonic functions. 
While the BPS solutions can be understood within the appropriate $\N=2$ 
truncation of $\N=8$ supergravity, the general non-BPS solutions require 
the whole field content of the theory.}
\end{abstract}

\end{titlepage}
\renewcommand{\thefootnote}{\arabic{footnote}}
\setcounter{footnote}{0}


\section{Introduction}

Extremal static black holes with identical charges do not interact. As 
first pointed out by Papapetrou and Majumdar, for every spherically symmetric 
extremal black hole of Maxwell--Einstein theory, one can arrive at associated 
multi-black hole solutions by means of linear `superposition', that is, by
substituting any harmonic function of the inverse radius in the potentials 
defining the stationary metric and the vector field \cite{Papapetrou,Majumdar}.
Physically, the stability (or `staticity') of these solutions is ensured by
the `no force' property, whereby the gravitational attraction between
any two charged black holes is exactly balanced  by their mutual
electromagnetic repulsion. On the other hand, considering Maxwell--Einstein 
theory as the bosonic sector of $\N=2$ supergravity, one can understand this 
property mathematically from the fact that the corresponding solutions 
are BPS, as they preserve one-half (that is, four) of the supersymmetry
charges. For general matter coupled $\N=2$ supergravity theories, there are 
multi-black hole solutions of BPS type for any black hole preserving four 
identical supersymmetry charges. Such solutions have been identified in 
a general way in \cite{Denef}, in the framework of the so-called
attractor mechanism \cite{attractors,attractors1}. 

However, the BPS multi-black hole solutions are not the most general 
solutions of this type. In this publication we characterise the 
general solutions of 
Papapetrou--Majumdar type for all models for which the scalar fields 
coordinatise a symmetric space. More precisely, our discussion applies to 
all models containing gravity coupled to abelian vector fields and scalar 
fields parametrising a symmetric Riemannian space $\G_4 / \H_4$ whose isometry 
group $\G_4$ acts faithfully on the vector fields. For these models, the 
stationary equations of motion reduce to the ones of a non-linear sigma model 
defined over a pseudo-Riemaniann symmetric space $\G/ \H^*$ coupled to 
three-dimensional Euclidean gravity  \cite{Maison,Maison1,nous}, with 
$\G$ being a simple Lie group, and $\H^*$ a {\em non-compact} real form of 
its maximal compact subgroup. The derivation of Papapetrou--Majumdar type 
solutions then reduces to certain algebraic conditions \cite{Clement} 
(see eqs.\ (\ref{tricom}) and (\ref{linenil}) of this paper). Using the 
formalism developed in \cite{nous}, and especially the explicit relation 
between extremal solutions and nilpotent orbits of the three-dimensional 
duality group $\G$, we are able to solve these algebraic equations 
in full generality.  A crucial feature here is that, unlike supergravity
duality groups in dimensions $D\geq 4$, the duality groups $\G$ of 
the three-dimensional theories also incorporate the Ehlers symmetry 
$SL(2,\mathbb{R})$, and thus the gravitational degrees of freedom. The importance of Lie algebra nilpotent orbits in the physics of extremal black holes was first pointed out in \cite{Boris}. The case of minimal supergravity in five dimensions has been studied in detail in this framework in \cite{Gaiotto}.

In contrast to previous studies of BPS solutions, the analysis of \cite{nous} 
is solely based on exploiting properties of the $\G$-Noether charge 
$\C$ and the associated BPS parameter $c$, which are defined
in eqs.~(\ref{C}) and (\ref{c}) below. As shown in \cite{nous}, the Noether 
charge satisfies a cubic or quintic characteristic equation, cf. 
(\ref{cubic}) and (\ref{quintic}), respectively. The latter determines 
the BPS parameter as a function of the physical charges of the 
four-dimensional theory, namely the complex gravitational charge $\w=m+in$ 
(with $m$ the mass and $n$ the Taub-NUT parameter) and the complex
(dyonic) electromagnetic charges $Z_{A}= q_A + ip_A$, {\it viz.}
\be\label{cwZ}
c = c(\w, Z_{A})
\ee

Solutions of Papapetrou--Majumdar type are superpositions of extremal 
black holes whose Noether charges satisfy the condition $c=0$. These 
solutions come in two categories. Those of the first category 
are \ft12 BPS solutions of an appropriate truncation to a theory that 
constitutes the bosonic sector of some $\N=2$ supergravity theory coupled 
to  $n_{\scriptscriptstyle V}$ vector multiplets.  The \ft12 BPS solutions 
of the latter theory define multi-black hole solutions depending on 
$2+2n_{\scriptscriptstyle V}$ harmonic functions 
($1+2n_{\scriptscriptstyle V}$ if one requires the solution to be 
asymptotically Minkowski). They define the most general solutions of 
Papapetrou--Majumdar type whenever the associated group $\H^*$ is 
non-semi-simple, \ie possesses a $U(1)$ factor. 
In this case, the BPS parameter can be written in the form
\be\label{cProd}
c^2 = \frac{ \scal{ |\w|^2 - |z_\un|^2 } 
\scal{Ê|\w|^2Ê- |z_\deux|^2 } }{|\w|^2} 
\ee
where $z_\un$ and $z_\deux$ generalise the central charges associated to 
the electromagnetic charges to theories which are not necessarily 
supersymmetric. It then follows that the extremality condition $c=0$ 
requires at least one of the charges to saturate the BPS bound 
$|z|=|\w|$. Consequently, the solution is \ft12 BPS within 
an appropriate $\N=2$ truncation.

When $\H^*$ is semi-simple, on the other hand, the BPS parameter $c$  
cannot be expressed in the simple form (\ref{cProd}), but is a more 
complicated non-rational function of the physical charges. In this case, the condition $c=0$ does not necessarily imply saturation 
of any BPS bound. This is
borne out by Papapetrou--Majumdar type solutions in the second category.
 As shown in \cite{nous} this can happen in particular 
for maximal $\N=8$ supergravity, or more generally for theories that can 
be obtained by dimensional reduction of five-dimensional supergravities
on a circle. As we will show, such theories with $n$ vector fields in 
four dimensions admit multi-black hole solutions depending on $1+n$ 
harmonic functions ($n$ for asymptotically Minkowski solutions) that 
do not define \ft12 BPS solutions of any $\N=2$ truncation of the theory. 

Because they rely essentially only on properties of the three-dimensional 
duality groups $\G$, our results in principle allow a general classification.
In order to bring out the main points as clearly as possible, we here 
restrict attention mostly to the example of  maximal $\N=8$ supergravity 
\cite{CJ}. The $\N = 4$ supergravities will be discussed in 
detail in a companion publication \cite{moi}. The extremal 
spherically symmetric black holes of maximal supergravity
can be understood within the simplified context of the so-called $STU$ 
$\N=2$ supergravity \cite{Ferrara}. From the point of view of the latter, 
there are three types of generic extremal spherically symmetric black holes: the 
\ft 12 BPS ones, the non-BPS ones for which the central charge vanishes 
on the horizon ($Z=0$), and the non-BPS ones for which the central charge 
does not vanish on the horizon ($Z \ne 0$) \cite{DallAgatta,Hotta,Gimon}. Both   
\ft 12 BPS and non-BPS black holes for which the central charge vanishes 
on the horizon correspond to \ft 18 BPS black holes within maximal 
supergravity. Within the $STU$ model such solutions are related by 
exchanging the vector fields belonging to the gravity multiplet with those 
in the matter multiplets, respectively. By contrast, 
the non-BPS solutions with $Z \ne 0$ are still non-BPS when embedded 
in maximal supergravity. As we will explain in this paper, such solutions 
are associated to theories that define compactification on a circle of 
theories defined in $4+1$ dimensions. For instance, pure gravity in five 
dimensions dimensionally reduced on a circle does admit extremal solutions, 
and it obviously defines a consistent truncation of any theory originating 
from five dimensions \cite{Gaiotto}. One then has a corresponding $\H^*$-orbit of extremal 
solutions, which for the $STU$ model corresponds to the non-BPS solutions 
with non-vanishing central charge on the horizon.

The plan of the paper is as follows. We first review the general 
formalism \cite{Maison,nous} and explain how the Papapetrou--Majumdar 
solutions can be read from the nilpotent orbits. Then we explain how one 
can understand the two categories of Papapetrou--Majumdar solutions within 
this formalism. We display in detail the spectrum of such solutions in 
maximal supergravity. In a subsequent publication by one of the authors 
\cite{moi}, the case of $\N=4$ supergravity coupled to $n$ vector multiplets 
will be discussed in detail. In particular, it will include explicit non-BPS 
multi-black hole solutions belonging to the second category (the would be 
non-BPS solutions with $Z \ne 0$ within the $STU$ model) depending on 
$6+n$ independent harmonic functions.

\section{Extremal solutions and nilpotent orbits}
Stationary solutions of Einstein theory of gravitation in four dimensions 
coupled to abelian vector fields and scalar fields parametrising a symmetric 
space $\G_4 / \H_4$ can be effectively obtained by solving the equations 
of motion of an associated non-linear sigma model coupled to gravity 
in {\em three} spatial dimensions (\ie with Euclidean signature). When the 
isometry group $\G_4$ acts on the abelian vector fields in a faithful 
representation $\mathfrak{l}_4$ to define a non-linear symmetry of the 
equations of motion, one thus arrives at a three-dimensional non-linear sigma 
which unifies all the physical bosonic degrees of freedom into a single
pseudo-Riemanian coset space $\G/\H^*$, with $\G$ a simple Lie group 
and $\H^*$ a non-compact real form of its maximal compact subgroup 
\cite{Maison,Maison1}. 

To derive the equations of motion for a coset 
representative $\V$ in  $\G/\H^*$, one decomposes the Maurer--Cartan form 
$\V^{-1} d \V$ into its coset and its $\mathfrak{h}^*$ components,
\be
\V^{-1} d\V = Q + P \quad, \qquad Q\equiv Q_\mu dx^\mu\in \mathfrak{h}^* 
  \;\; , \;\;\; 
 P\equiv P_\mu dx^\mu\in \mathfrak{g}\ominus\mathfrak{h}^* 
\ee
where $\mu,\nu,\dots=1,2,3$. The equation of motion for the scalars reads
\be
d   \star \V  P  \V^{-1}Ê = 0
\label{EinsteinE}
\ee
where $\star$ is the Hodge star operator associated to the three-dimensional 
Riemannian metric $g_{\mu\nu}$.  The Einstein equation is
\be\label{Einstein}
R_{\mu\nu} = \trace P_\mu P_\nu 
\ee
Because in three dimensions the Riemann tensor is entirely determined
by the Ricci tensor, the three-dimensional metric $g_{\mu\nu}$ is flat if the right hand side of
(\ref{Einstein}) vanishes.

Black hole solutions of the four-dimensional theory 
correspond to instantons of the associated three-dimensional non-linear sigma model 
over $\G/\H^*$. The spherically symmetric black holes (including the asymptotically 
Taub--NUT ones) are entirely characterised by their $\G$-Noether charge
\be 
\mathcal{Q} \equiv  \frac{1}{4\pi} \int_{\partial V}  \star \V P \V^{-1}  
\ee
and the asymptotic value of the scalars fields $\V_\zero \in \G_4 \subset \G$ 
at spatial infinity \cite{Maison}. Here it will be more convenient to characterise the solutions 
in term of a modified conserved charge $\C$ obtained by rotating 
$\mathcal{Q}$ back into the coset
\be\label{C} 
\C \equiv   {\V_\zero}^{-1}    \mathcal{Q}  \, \V_\zero  
\in \mathfrak{g} \ominus \mathfrak{h}^* 
\ee
which we will call the `Noether charge' for simplicity (this designation being 
unambiguous since we will never refer to $\mathcal{Q}$ itself). 

Any spherically symmetric black hole then admits the following  
$\G/\H^*$ representative \cite{Breit}
\be\label{V} 
\V = \V_\zero \, \exp \left( \frac{1}{2 c} \ln \Scal{ \frac{ r-c}{r+c} }\, 
\C \right)  
\ee
for some matrix $\C$, where $r$ is the Weyl radius, associated to the 
three-dimensional Riemannian metric~\footnote{The Weyl radius $r$
  is related to the standard radial coordinate $\tilde{r}$ via
  $\tilde{r} = r+m$ such that e.g. the horizon of the Schwarzschild 
  solution corresponds to $r=m$ (see e.g. \cite{Book}). Unlike the usual 
  Schwarzschild coordinates the Weyl coordinates are duality invariant, 
  as follows from the duality invariance of the BPS parameter $c$ and 
  the split of the four-metric in (\ref{4dmetric}) below.} 
\be 
g_{\mu\nu} dx^\mu dx^\nu  = dr^2 + 
(Êr^2 - c^2 ) \scal{ d\theta^2 + \sin^2 \theta d\varphi^2 } \label{sphere} 
\ee
The BPS parameter $c$ is defined in terms of the Noether charge as 
\be\label{c}
c^2 \equiv \frac{1}{k} \,\trace \C^2\, ,
\ee 
Here $k\equiv  \trace {\bf h}^2$ is a positive integer associated to the 
group $\G$ and the following five-graded decomposition of its Lie algebra 
$\g$ with respect to its subalgebra $\gl_1 \equiv \mathbb{R} {\bf h}$ 
\be 
\g \cong   {\bf 1}^{\ord{-2}} 
\oplus \mathfrak{l}_4^{\ord{-1}} \oplus \left(\gl_1 \oplus 
\mathfrak{g}_4^\ord{0} \right) \oplus \mathfrak{l}_4^\ord{1} 
\oplus {\bf 1}^\ord{2} \label{five}
\ee
This decomposition shows explicitly how the duality group $\G_4$ of the four-dimensional
theory and its representation $\mathfrak{l}_4$ are embedded in $\G$.

Given a solution of the three-dimensional equations of motion one
can reconstruct the solution in four dimensions as follows. In order 
to read off the Kaluza--Klein ansatz for the four-dimensional metric and 
the abelian vector fields from the above  `data', one must first rotate the coset 
representative $\V$ into a parabolic (triangular) gauge $\V \in \mathfrak{P}$ 
where $\P$ is the maximal parabolic subgroup of $\G$  defined from the 
five-graded decomposition (\ref{five}) with Lie algebra $\mathfrak{p}$ 
\be \mathfrak{p} \cong  \left( \gl_1 \oplus \mathfrak{g}_4^\ord{0}\right)  
\oplus \mathfrak{l}_4^\ord{1} 
\oplus {\bf 1}^\ord{2} \label{parabolic}\ee
Let us designate by $\ddagger$ the involution that defines 
$\mathfrak{h}^* \subset \mathfrak{g}$, such that ${\bf x} - {\bf x}^\ddagger 
\in \mathfrak{h}^*$. We write $\bf h$ for the $\gl_1$ generator, and 
$\bf e$ and ${\bf f} = {\bf e}^\ddagger$ for the grade $2$ and $-2$
generators, respectively. The representation $\mathfrak{l}_4$ always 
admits a $\G_4$ invariant symplectic form (which defines the Dirac 
quantisation condition) such that two generators $\bf x$ and $\bf y$ 
of $\mathfrak{l}_4^\ord{1}$ commute to 
\be 
[ {\bf x} , {\bf y} ] = ( {\rm x }, {\rm y} )_{\mathfrak{l}_4}  {\bf e} 
\ee
Defining $\boldsymbol{\Phi} $ as the electromagnetic scalars valued in $\mathfrak{l}_4^\ord{1}$ and $v \in \G_4 / \H_4$ the four-dimensional scalars fields, one writes down the coset representative $\V$ in the parabolic gauge as
\be 
\V = \exp\scal{ÊB {\bf e} +  \boldsymbol{\Phi} } \, 
\exp\left( \frac{1}{2}Ê\ln H \, {\bf h}\right) \, v  \label{Pgauge} 
\ee
A straightforward computation gives
\begin{multline}  \label{P}
P = \frac{1}{2}ÊH^{-1}  d H \, {\bf h} + \frac{1}{2} H^{-1} \Scal{ d B - \frac{1}{2} \scal{Ê \Phi , d \Phi }_{\mathfrak{l_4}} } ( {\bf e} + {\bf f})  \\* + \frac{1}{2}ÊH^{-\frac{1}{2}} L_4(v^{-1}) Êd  ( \boldsymbol{\Phi} + \boldsymbol{\Phi}^\ddagger )   + \scal{Êv^{-1} d v }_{|\mathfrak{g}_4 \ominus \mathfrak{h}_4}  \end{multline}
where $L_4$ is the $\mathfrak{l}_4$ representation homomorphism. Substituting
these expressions into the equation of motion (\ref{EinsteinE}) it follows that
the axion field $B$ defines the Kaluza--Klein vector $\hat{B} $ through 
\be  
H^{-2} \star  \Scal{ d B - \frac{1}{2} \scal{Ê \Phi , d \Phi }_{\mathfrak{l_4}} } = d \hat{B} \label{dual} 
\ee
such that the four-dimensional metric reads
\be\label{4dmetric}
ds^2 = - H \scal{Êdt + \hat{B}_\mu dx^\mu }^2 + 
H^{-1} \, g_{\mu\nu} dx^\mu dx^\nu 
\ee
In order to reconstruct the four-dimensional vector fields 
\be 
\sqrt{2\pi G}Ê\mathcal{A} = U \scal{Êdt + \hat{B}_\mu dx^\mu} 
+ \hat{A}_\mu dx^\mu 
\ee
from (\ref{P}) one must choose a Lagrangian subspace of $\mathfrak{l}_4$ 
with respect to the symplectic form of $\G_4$, such that $\Phi$ splits into 
$U \oplus A$. After the redefinition $B^\prime \equiv B + \frac{1}{2}  
( U, A )_{\mathfrak{l}_4}$, the fields $A$ only enter $P$ linearly through 
their differential $dA$, and can then be dualised to vectors $\hat{A}$ 
according to their equations of motion.  

Let us consider a general non-rotating asymptotically flat solution which 
does not carry any naked singularity, by which we mean that any singularity 
is covered by an horizon which is identified with its corresponding Killing 
horizon (no ergosphere). Such solutions are spherically symmetric black holes \cite{Maison}.
From the point of view of the three-dimensional Riemannian space $V$, the 
horizon $\mathscr{H}$ corresponds to the single point $r=c$ on $V$ (cf.
(\ref{sphere})) with an instanton-like singularity. Because the space-time 
volume element $H^{-1} \sqrt{g}$ is regular on the horizon of a non-extremal 
black hole,  while $H$ goes to zero there, the  three-dimensional volume 
element $\sqrt{g}$  vanishes on the horizon. It follows that {\em all} 
components of $\V$ must tend to the same value as $r\rightarrow c$ and 
are thus constant on $\mathscr{H}$. Using the fact that the 
dependency on $\hat{B}$ can be neglected on $\mathscr{H}$, the expression 
of the horizon area $A_\mathscr{H}$ in the coordinates (\ref{sphere}) 
\be 
A_\mathscr{H} \equiv \int_{\mathscr{H}} (  {r}^2 - {c}^2 ) H^{-1} 
\sin \theta d \theta \wedge d \varphi 
\ee
shows that for $A_\mathscr{H} >0$, $H$ behaves like 
\be H = \frac{4 \pi}{A_\mathscr{H}} (  {r}^2 - {c}^2 ) + \mathcal{O} \scal{Ê (  {r}^2 - {c}^2 )^\frac{3}{2}} \ee
near $\mathscr{H}$. Similarly, the surface gravity $\upkappa$ can be 
computed from the Killing vector $\xi^M \partial_M \equiv \partial_t$ as
\be
\upkappa^2 \equiv -\frac{1}{2}\lim_{r\rightarrow c}\,\nabla^M \xi^N \nabla_M \xi_N
 =  \frac{1}{4}  \left[\partial^\mu H \partial_\mu H - 
 H^4 \partial^\mu \hat{B}^\nu  (  
\partial_\mu \hat{B}_\nu -    \partial_\nu \hat{B}_\mu )\right]_{\mathscr{H}} 
\ee 
Exploiting the fact that $\hat{B}$ is regular on the horizon we thus obtain
\be
\upkappa =
\frac{1}{2} \partial_r H \Big|_{\mathscr{H}}=  \frac{ 4 \pi c }{ A_\mathscr{H} } 
\ee
Similarly, using the behaviour of the fields near the horizon, one computes 
that the charge associated to the horizon is given by 
\be  
\C \equiv \frac{1}{4\pi}   {\V_\zero}^{-1}    
\int_{\mathscr{H}}  \star \V P \V^{-1}  \, \V_\zero  =   
\lim_{r\rightarrow c} \Big( ( {r}^2 - {c}^2 )   {\V_\zero}^{-1}   
\V P_r \V^{-1} {\V_\zero} \Big)   
\ee
It follows that 
\be \trace {\C}^2 =  \lim_{r \rightarrow c}    
\Big( ( {r}^2 - {c}^2 )^2  \, \trace {P_r}^2 \Big)
\ee
Using the behaviour of $H$ near the horizon, the equation of motion 
(\ref{dual}) and the assumption that  all the fields $\hat{B}$, $\Phi$ 
and $v$ are regular on the horizon, we arrive at 
\be  
(  {r}^2 - {c}^2 ) \,  P_r = c \, {\bf h} + 
\mathcal{O}\scal{  \sqrt{ {r}^2 - {c}^2 }} 
\ee
recovering the formula (\ref{c}) as expected. Therefore, we obtain 
that the product of the horizon area and the surface gravity of a 
non-rotating black hole is determined by the Killing norm of the $\G$-Noether charge as 
\be A_\mathscr{H}  \upkappa = 4\pi \, c \ee
As a result, non-rotating extremal solutions, for which $\upkappa =0$, carry a 
Noether charge satisfying $\trace {\C}^2  = 0$ 
(since $\upkappa$ is identified with the temperature of the black hole
in the thermodynamic interpretation, it follows that all such solutions 
are characterized by zero temperature). Therefore, from (\ref{sphere}) we
see that  spherically symmetric extremal black holes solutions can be viewed as
{\em  instantons over flat three-dimensional space}, with the {\em Euclidean} 
three-dimensional metric $g_{\mu\nu} = \delta_{\mu\nu}$. For these solutions the coset 
representative $\V$ takes the form
\be\label{VC}
\V =  \V_\zero \, \exp \left(  - \frac{1}{r}  \, \C \right)  
\ee
with a Noether charge satisfying $\trace \C^2  = 0$. Inspection of the equations 
of motion shows that $\frac{1}{r}$ can be replaced by any solution of 
the three-dimensional Laplace equation $\Delta \Ha = 0$. Such solutions 
were first discovered by Papapetrou and Majumdar in Maxwell--Einstein 
theory \cite{Papapetrou,Majumdar}. The main purpose of this paper is to study
these solutions in a systematic manner by exploiting the maximal duality 
symmetry $\G$ acting on them.

\section{Solutions of Papapetrou--Majumdar type}

To obtain the solutions of Papapetrou--Majumdar type, we replace the 
formula (\ref{VC}) by the general Ansatz for the coset representative $\V$
\be 
\V (x)=   \V_\zero \, \exp\left(- \sum_\n \Ha^\n (x) \C_\n \right) \label{Ansatz} 
\ee
with Lie algebra elements $\C_\n \in  \mathfrak{g} \ominus \mathfrak{h}^*$,
and where the functions $\Ha^\n(x)$ are to be determined from the equations
of motion. It is straightforward to see that $P$ depends 
linearly on the functions $\Ha^\n$ if and only if \cite{Clement}
\be [Ê\C_\m , [ \C_\n , \C_\p ] ] = 0 \label{tricom}\ee
Under this assumption, one obtains that 
\be P = \sum_\n d \Ha^\n \C_\n \hspace{20mm} Q =
\frac{1}{2} \sum_{\n\m}  \Ha^\n d \Ha^\m  [ \C_\n , \C_\m ] \ee
and the equations of motion reduce to
\be R_{\mu\nu}  = \sum_{\m\n} \partial_\mu \Ha^\m \partial_\nu \Ha^\n \, \trace \C_\m \C_\n  \hspace{20mm}  d \star d \Ha^\n = 0 \ee
For the equations of motion to be linear in $\Ha^\n$, one must require 
in addition that 
\be \trace  \C_\m \C_\n = 0 \label{linenil} \ee
Any set of matrices satisfying (\ref{tricom}) and (\ref{linenil}) yields a solution 
of the theory with a flat three-dimensional metric, if the $\Ha^\n$ are arbitrary harmonic functions. 
Therefore the problem of solving the three-dimensional field equations can 
be reduced to solving the algebraic equations (\ref{tricom}) and (\ref{linenil}). 
The main new insight  from the present analysis (and from \cite{nous}) is that
this problem, in turn, {\em can be reduced to the construction and classification 
of the nilpotent orbits of $\g$ via their corresponding normal triplets (see below) 
and the associated graded decompositions of $\g$}, as we shall now explain.

Any element of a Lie algebra $\g$ can be written as the sum of a 
diagonalisable element and a nilpotent element. There is thus no loss 
of generality in assuming that each $\C_\n$ in the general ansatz
(\ref{Ansatz}) is either nilpotent or diagonalisable. Two elements $\C_\un$ 
and $\C_\deux$ satisfying equation (\ref{tricom}) either commute, or 
generate a Heisenberg subalgebra of $\g$. From the representation theory 
of Heisenberg algebras it follows that any linear combination of 
$\C_\un$ and $\C_\deux$ is nilpotent in the latter case. Therefore 
the matrices $\C_\n$ are either diagonalisable elements commuting with 
all the others, or nilpotent elements such that any linear combination 
of them is likewise nilpotent.

The set of nilpotent 
elements of $\g$ is known to decompose into finitely many $\G$-orbits, the 
so-called nilpotent orbits of $\g$  \cite{coadjoint}. For a given nilpotent 
orbit, one can associate to a given representative ${\bf E} \in \mathfrak{g}$ 
of the orbit, an $\sl_2$ triplet $({\bf H},{\bf E},{\bf F})$ such 
that \cite{coadjoint}
\be 
[ {\bf H} ,{\bf  E} ] = 2 {\bf E} \hspace{10mm} [{\bf H},{\bf F}] 
= -2 {\bf F} \hspace{10mm} [{\bf E},{\bf F}] = {\bf H} \label{triplet} 
\ee
Decomposing the Lie algebra with respect with the eigenvalues of $\bf{H}$, 
one gets a $(2n+1)$-graded decomposition of $\g$ (where
$\mathfrak{gl}_1 \equiv \mathbb{R} {\bf{H}}$)
\be\label{grad} 
\g \cong \g^\ord{-n} \oplus \cdots \oplus \g^\ord{-2} \oplus \g^\ord{-1} 
\oplus\left( Ê\gl_1 \oplus \g^\ord{0} \right) \oplus \g^\ord{1} \oplus 
\g^\ord{2} \oplus \cdots \oplus \g^\ord{n}
\ee  
such that $\g^\ord{0}$ is a reductive Lie algebra (\ie the direct sum of a 
semi-simple and an  abelian  algebra) which contains a Cartan subalgebra 
of $\g$. From equation (\ref{triplet}), it follows that the nilpotent 
element $\bf{E}$ lies in $\g^\ord{2}$, and the nilpotency degree 
of ${\bf E}$ in the adjoint representation~\footnote{ 
  Which is defined to be the smallest integer $k$ such that
  $(\ad_{\bf E})^k({\bf x}) = 0$ for all ${\bf x}\in\g$.}
 is the lowest integer strictly greater 
than $n$. The nilpotency degree then follows directly from the 
grading and is not altered by the detailed commutation rules of the Lie 
algebra in this graded decomposition. For a general nilpotent orbit, $n$ 
can be pretty large, but if we consider a nilpotent orbits for which 
$n\le 5$, it follows from the graded decomposition that any 
element $\C_\n \in\bigoplus_{p=2}^n \mathfrak{g}^\ord{p}$ satisfies 
both equations (\ref{tricom}) and (\ref{linenil}). There are four 
distinguished cases, namely
\bea
{ \ad_{\bf E} }^6 = 0\hspace{10mm}    &\Rightarrow& \hspace{10mm}   n = 5\CR 
{ \ad_{\bf E} }^5 = 0\hspace{10mm}    &\Rightarrow& \hspace{10mm}   n = 4\CR 
{ \ad_{\bf E} }^4 = 0 \hspace{10mm}   &\Rightarrow& \hspace{10mm}   n = 3\CR
{ \ad_{\bf E} }^3 = 0 \hspace{10mm}   &\Rightarrow& \hspace{10mm}   n = 2
\label{gradlow} \eea
All of the above relations also hold for the complexification $\G_{\mathds{C}}$
of the duality group $\G$ and its associated complex Lie algebra $\g_{\mathds{C}}$.
What is important is that the complex nilpotent orbits in $\g_{\mathds{C}}$
are {\em uniquely} determined by the nilpotency degree (in several representations), and equivalently by the 
graded decomposition (\ref{grad}). When descending from the complex to
the real Lie algebra, the complex nilpotent orbit may decompose into several
real orbits, which are the ones relevant for black hole solutions.

The isotropy subalgebra of a representative ${\bf E}$ of a given orbit is a subalgebra of $\bigoplus_{p=0}^n \g^\ord{p}$. Any solution to equation (\ref{tricom}) can thus be associated to a given $2n+1$ graded decomposition (\ref{grad}) with $n\le 5$, such that the nilpotents elements lie in $\bigoplus_{p=2}^n \mathfrak{g}^\ord{p}$ and the diagonalisable elements commute with all the others and lye in $\g^\ord{0}$. Equation (\ref{linenil}) is trivially satisfied for the nilpotent elements, but strongly constrains the diagonalisable elements of $\g^\ord{0}$.

More specifically, in black hole physics, we are interested in elements 
which lie in $\g \ominus \h^*$ in order for the coset representative $\V$ to be in 
the symmetric gauge. In fact, one can show that there is no loss of 
generality by doing so, because any more general coset element can be 
rotated back to this form by a right $\H^*$ gauge transformation. When 
the $\G$-orbit of the nilpotent element ${\bf E}$ has a non-trivial 
intersection with $\g \ominus \h^*$, 
the triplet can be chosen in such a way that both ${\bf E}$ and ${\bf F}$ 
lie in $\g \ominus \h^*$, and such that ${\bf H}$ lies in $\h^*$. 
It follows then that $\h^*$ decomposes in a similar way as 
\be 
\h^* \cong \h^\ord{-n} \oplus  \cdots \oplus \h^\ord{-2} \oplus \h^\ord{-1} 
\oplus\left( Ê\gl_1 \oplus \h^\ord{0} \right) \oplus \h^\ord{1} \oplus 
\h^\ord{2} \oplus \cdots \oplus \h^\ord{n} 
\ee 
such that $\h^\ord{0}$ is a reductive Lie algebra which contains a Cartan 
subalgebra of $\h^*$, and ${\bf E} \in \g^\ord{2} \ominus \h^\ord{2}$. The 
elements of $ \g^\ord{p} \ominus \h^\ord{p}  $ (for $p\ge 2$) generate 
a nilpotent Lie algebra $\mathfrak{n}^\ord{p}$, which decomposes as 
(for $n\le 5$)
\be \mathfrak{n}^\ord{p} \cong ( \g^\ord{p} \ominus \h^\ord{p} ) \oplus  \h^\ord{2p}  \label{ndec} \ee
Any set of charge matrices $\C_\n$ lying in $\bigoplus_{p=2}^n ( \g^\ord{p} 
\ominus \h^\ord{p})$ therefore satisfies the commutation relations 
(\ref{commut}) and defines a multi-black hole solution via the 
Ansatz (\ref{Ansatz}). In the case of the graded decomposition (\ref{five}), 
which will be the one of interest as we are going to see, the isotropy 
subgroups of grade zero $\mathfrak{J}^\ord{0}_\n\subset \G$ and 
$\mathfrak{K}^\ord{0}_\n \subset \H^*$ of a generic element ${\bf E} 
\in \mathfrak{l}_4^\ord{2} \ominus \h^\ord{2}$ coincide with the isotropy 
subgroups of $\G_4$ and $\H_4$ of the corresponding electromagnetic charges, 
respectively. These subgroups also define the so-called moduli spaces of 
black hole attractors as  $ \mathfrak{J}^\ord{0}_\n / \mathfrak{K}^\ord{0}_\n 
\subset \G_4 / \H_4$  \cite{Ferrara}. The Cartan norm is thus positive 
definite on $\mathfrak{j}^\ord{0}_\n \ominus \mathfrak{k}^\ord{0}_\n$ 
and there is no diagonalisable element to consider in the solutions 
of (\ref{tricom}, \ref{linenil}).

Let us next consider a more general Ansatz than (\ref{Ansatz}) including 
Lie algebra elements $\mathscr{A}_\s \in \bigoplus_{p=2}^n 
\mathfrak{h}^\ord{n}$. From the grading and the Campbell--Hausdorff formula 
it follows that 
\begin{multline}  \V =  \V_\zero  \, \exp\left(- \sum_\n \Ha^\n \C_\n  - \sum_\s \mathcal{K}^\s \mathscr{A}_\s  \right) \\*  =   \V_\zero  \, \exp\left(- \sum_\n \Ha^\n \C_\n  + \frac{1}{2}  \sum_{\s\, \n}  \mathcal{K}^\s \Ha^\n [ \mathscr{A}_\s  , \C_\n ]   \right)  \, \exp\left(- \sum_\s \mathcal{K}^\s \mathscr{A}_\s  \right) \end{multline}
such that this Ansatz is in fact equivalent to (\ref{Ansatz}) up to a right $\H^*$ gauge transformation.\footnote{Note that  $[ \mathscr{A}_\s  , \C_\n ] \in \bigoplus_{p=4}^n ( \g^\ord{p} \ominus \h^\ord{p} )$.}
 
The solutions obtained by the above construction may in principle
exhibit naked singularities. Quite generally, the latter can be of two types. 
For solutions not obeying any `no force' property, there are usually rod-like
singularities which appear in static multi-black hole solutions in order 
to balance the gravitational pull between any two black holes,
and which manifest themselves as string-like singularities of the Riemannian 
metric $g_{\mu\nu}$. For example, in the axisymmetric case,\footnote{In which 
  case exact solutions with several black holes that do not share any 
  no force property are known explicitly \cite{Book}.} they appear 
in Weyl coordinates 
\be 
g_{\mu\nu} dx^\mu dx^\nu = e^{2 \sigma}Ê\scal{Êdz^2 + d \rho^2 } 
+ \rho^2 d \varphi^2 \ee
if the function $\sigma(\rho,z)$ does not vanish in the limit 
$\rho \rightarrow 0$ on the axis in between two interacting black holes.
This type of singularity cannot occur for the solutions built on the
conditions (\ref{tricom}) and (\ref{linenil}) because the three-dimensional 
metric is then flat and regular everywhere by (\ref{Einstein}). Consequently,
the singularities of the four-dimensional theory solely originate
from the singularities of the scalar fields which are located at the poles 
of the harmonic functions $\Ha^\n$. Naked singularities of such solutions are 
thus entirely due to the individual black holes themselves, and can be 
avoided by choosing each charge $\C_\n$ so that the individual 
black holes have their singularities covered by  horizons.  
This way one makes sure that each pole of the harmonic functions 
corresponds to an horizon in four dimensions.

It is commonly assumed that all the spherically symmetric extremal black 
holes without naked singularities correspond to particular limits of 
non-extremal spherically symmetric black holes which do not carry any 
naked singularity themselves. Within the class of model discussed in 
this paper, the asymptotically flat regular non-extremal solutions are 
all in the $\H^*$-orbit of 
uncharged Kerr-solutions with scalar fields having any constant 
value \cite{Maison}. It follows that the corresponding Noether charge 
$\C$ satisfies a cubic characteristic equation \cite{nous} 
\be \C^3 = c^2 \, \C \label{cubic}  \ee
in the fundamental representation (which is the spinor representation of 
its double-cover when $\G$ is an orthogonal group), save for two particular 
cases for which $\G$ is a non-compact real form of $E_{8}$, corresponding
to maximal $\N=8$ supergravity and the `magic' $\N=2$ supergravity associated to the octonions \cite{magic}. In these 
two cases the characteristic equation is quintic \cite{nous}, and must be 
satisfied in the ${\bf 3875}$ representation that appears in the symmetric 
product of two copies of the adjoint representation,
\be \C^5 = 5 \, c^2 \C^3 - 4 c^4  \, \C \label{quintic} \ee
It follows that regular extremal solutions (for which $c=0$) carry a nilpotent 
Noether charge $\C$, which vanishes at the third power in the fundamental 
representation, \ie $\C^3 = 0$, in the generic case, or at the fifth 
power in the ${\bf 3875}$ representation of $E_8$ when $\G$ is a real 
form of $E_8$ \cite{nous}. So as a first consequence, one can restrict attention to nilpotent elements $\C_\n$ in considering regular multi-black holes, because any diagonalisable elements of vanishing Killing norm would lead to naked singularities.

The nilpotency conditions are preserved by the action of the complexified 
group $\G_\mathds{C}$, and they determine one single complex nilpotent 
orbit with  representative ${\bf E}_\cinq$. As already mentioned the complex
orbit may decompose into several real nilpotent orbits when one descends 
to the real group $\G\subset\G_{\mathds{C}}$. Acting with the complex
Lie group $\G_{\mathds{C}}$ on the representative element $\bf{E}_\cinq$ 
we obtain a dense open subset within the complex variety defined by 
the nilpotency condition  \cite{coadjoint}. Therefore, any solution $\C$ of the characteristic 
equation (\ref{cubic}) or (\ref{quintic}) lies in the closure of this 
complex orbit inside the variety of nilpotent elements of $\g_\mathds{C}$.
Since, in addition we require this orbit to be {\em real} and to have
a non-trivial intersection with $\g \ominus \h^* $, we have
\be
\C \in \overline{\G_\mathds{C} \cdot {\bf E}_\cinq} 
\cap \g \ominus \h^* 
\ee
It turns out that the $\sl_2(\mathds{C})$ triplet associated to 
${\bf E}_\cinq$ is such that ${\bf H}_\cinq = [Ê{\bf E}_\cinq , 
{\bf F}_\cinq ]$ satisfies the same characteristic 
equation as $2{\bf h}$, and the graded decomposition of $\g_\mathds{C}$ 
associated to such nilpotent element is therefore the (complexified) 
five-graded decomposition associated to the dimensional reduction 
(\ref{five}). The five-graded decomposition of $\g_\mathds{C}$ with respect
to ${\bf H}_\cinq$ corresponds 
to a unique five-graded decomposition of $\g$,\footnote{Although it can 
  be degenerate, as for example for $\e_{6(-14)}$ and $\e_{7(-25)}$. 
  Nevertheless the orbit we are interested in is unique in these cases.}
\be \g \cong   {\bf 1}^{\ord{-4}} 
\oplus \mathfrak{l}_4^{\ord{-2}} \oplus \left( \gl_1 \oplus 
\mathfrak{g}_4^\ord{0} \right) \oplus \mathfrak{l}_4^\ord{2} 
\oplus {\bf 1}^\ord{4} \label{fiveDbl} \ee
In general, the elements ${\bf E}$ and ${\bf H}$ of a normal triplet do not satisfy 
the same characteristic equation, and this property is very particular 
to ${\bf h}$ and its five-graded decomposition. In fact, this five-graded 
decomposition characterises both the minimal semi-simple $\G_\mathds{C}$-orbits
of $\g_\mathds{C}$, \ie $\G_\mathds{C} \cdot {\bf h}Ê\cong \G_\mathds{C} / 
( \mathds{C}^{\scriptscriptstyle \times } \times \G_{4\, \mathds{C}})$, and 
the minimal nilpotent $\G_\mathds{C}$-orbit  of $\g_\mathds{C}$, \ie 
$\G_\mathds{C}  \cdot {\bf e} \cong \G_\mathds{C} / ( \G_{4\, \mathds{C}} 
\ltimes ( \mathfrak{l}_{4\, \mathds{C}}^\ord{1} \oplus \mathds{C}^\ord{2}))$.
   
In order to correspond to a regular solution, the Noether charge $\C$ must 
lie inside the closure of the $\H^*$-orbits inside $\G_\mathds{C} \cdot 
{\bf E}_\cinq \cap \g \ominus \h^* $,  whose isotropy subgroups are contracted 
forms of $\H_4$ \cite{nous}, since the latter is the isotropy subgroup of 
the Kerr solutions. $\H_4$ being compact, the semi-simple component of the isotropy subgroup of $\H^*$ of  such charges is also compact. The most general solutions of Papapetrou--Majumdar type 
correspond to linear combination of such nilpotent charges which vanish 
at the sixth power in the adjoint representation. When considering a nilpotent element within the graded decomposition associated to a higher order orbit, its isotropy subgroup generally involves components of negative grading such that the semi-simple component of its isotropy subgroup appears itself through a graded decomposition embedded inside the graded decomposition of $\h^*$ associated to the higher order orbit. It follows that the latter isotropy subgroup cannot be a contracted form of $\H_4$. The relevant linear combinations of charges therefore all lye in the closure of the $\H^*$ orbits of $\G_\mathds{C} \cdot  {\bf E}_\cinq \cap \g \ominus \h^* $. We will prove explicitly in \cite{moi} that this is indeed the case within $\N=4$ supergravity coupled to $n$ vector multiplets. The only nilpotent orbits we have to consider in order to obtain 
the most general solutions of Papapetrou--Majumdar type are thus the {\em real 
nilpotent orbits of $\g$} of dimension $2 \dim[\mathfrak{l}_4] + 2$ 
associated to the five-graded decomposition (\ref{fiveDbl}), or more 
precisely, the $\dim[\mathfrak{l}_4] + 1$ dimensional $\H^*$-orbits of 
nilpotent elements inside $\g\ominus \h^*$ associated to this five-graded 
decomposition, such that the $\gl_1$ subalgebra lies in $\h^*$. 
In general, there are different inequivalent embeddings of $\h^* \subset \g$ 
consistent with the graded decomposition (\ref{fiveDbl}),
\be \h^* \cong  {\h}^{\ord{-4}} 
\oplus \mathfrak{h}^{\ord{-2}} \oplus \gl_1 \oplus 
\mathfrak{h}^\ord{0}  \oplus \h^\ord{2} 
\oplus \h^\ord{4} \label{Hfive} \ee
which correspond to the different $\H^*$-orbits inside $\G_\mathds{C} \cdot {\bf E}_\cinq \cap \g \ominus \h^* $. There are inequivalent Papapetrou--Majumdar solutions associated to each of those graded decompositions, whose number of independent harmonic functions are $\dim[ \mathfrak{l}_4^\ord{2} \ominus \h^\ord{2} ] + 1 -  \dim[\h^\ord{4}] $. 

\section{Supersymmetric solutions}

When the theory coincides with the bosonic sector of a supergravity theory, 
it is well known that BPS black holes preserving identical supersymmetry 
charges define multi-particle solutions of Papapetrou--Majumdar type. 
We are now going to see how this can be understood in the general 
framework developed in the foregoing section. To this aim we recall a
main result of \cite{nous}, namely the fact that the Noether charge
admits a dual description, either as a {\em matrix} $\C$ valued in 
$\g \ominus \h^*$ as in (\ref{C}), or alternatively, and equivalently, 
as a {\em state} $|\C\rangle$ in a fermionic Fock space, transforming 
in the same representation as the coset matrix. In the latter description
the BPS parameter (\ref{c}) is simply given by the indefinite `norm' 
of this state, to wit
\be\label{NormC}
c^2 = \langle\C|\C\rangle
\ee

For $\N$-extended supergravity, the non-compact group $\H^*$ is the 
product of the chiral component\footnote{By chiral component of $Spin^*(2\N)$,
  we mean its subgroup $Spin^*(2\N) / \ker(S_+)$ that acts faithfully in 
  the chiral Weyl spinor representation $S_+$, \ie $SL(2,\mathds{R}) 
  \subset Spin^*(4)$, $SO(2,6) \subset Spin^*(8)$, $Spin^*(16) / \mathds{Z}_2 
  \subset Spin^*(16)$ and $Spin^*(2\N)$ itself for odd $\N$.} of the 
R-symmetry group $Spin^*(2\N)$ and a group $\H^*_0$ depending on the 
matter content of the theory. The supersymmetry parameters are $SU(2)$ 
spinors valued in the pseudo-real vector representation of $SO^*(2\N)$ 
which satisfy an $SO^*(2\N)$-Majorana condition associated to the 
pseudo-anti-involutions of $SU(2)$ and $SO^*(2\N)$. Using 
$U(\N)\subset SO^*(2\N)$ covariant notations, the spinor parameters 
are complex $SU(2)$ spinors $\epsilon_\alpha^i$ valued in the fundamental 
representation of $U(\N)$, with complex conjugates $\epsilon_i^\alpha$. 
In order for the solution to be supersymmetric, the corresponding Noether 
charge state must satisfy the `Dirac equation' \cite{nous} 
\be\label{Dirac1} 
\scal{Ê\epsilon_\alpha^i a_i + 
\varepsilon_{\alpha\beta} \epsilon^\beta_i a^i } | \C \rangle = 0 
\ee
where $a^i$ and $ a_j$ (for $i,j,\dots = 1,...,\N$) are the fermionic 
oscillators from which the spinor representations of $SO^*(2\N)$ are built 
(see appendix~B of \cite{nous} for details). The condition (\ref{Dirac1})
is derived by considering the supersymmetry variation of the dilatino 
fields in the asymptotic region. Observe that it contains more detailed 
information than the gravitino variation, which simply yields the 
necessary condition $c=0$ for Killing spinors to exist (the latter 
statement also follows from (\ref{Dirac1}) as an integrability 
condition). In searching for extremal solutions, we can thus replace the 
problem of solving the cubic or quintic nilpotency conditions
for the charge matrix $\C$ by the simpler linear condition (\ref{Dirac1}) 
for the associated charge state $|\C\rangle$ \cite{nous}.

Solutions of (\ref{Dirac1}) are characterised by the number of
supersymmetries  and by their isotropy groups. Generally, the latter 
are always {\em parabolic subgroups} of $\H^*$  because the 
charge state $|\C\rangle$ is a {\em null vector}  (\ie a zero norm state) 
for $c=0$ by (\ref{NormC}). In \cite{nous} we showed that solving
the characteristic equations (\ref{cubic}) or (\ref{quintic}) for regular solutions
gives rise to a stratified moduli space
\be\label{Strata}
\M = \M_0 \cup \M_1 \cup \dots \cup \M_n
\ee
where the main stratum $\M_0$ corresponds to non-extremal solutions
(that is, $c^2 >0$), while the remaining strata correspond to solutions
with $c=0$ and contain the supersymmetric or BPS solutions as a subspace.
When the solution is left invariant by $4n$  supersymmetry generators, 
one speaks of an `\ft n\N - BPS solution'. In this case a basis of 
supersymmetry generators can be chosen with spinor parameters 
$\epsilon_\alpha^A$, with $A$ running from $1$ to $2n$, which
satisfy the reality condition ({\it alias} symplectic Majorana condition)
\be 
\epsilon_\alpha^A + \varepsilon_{\alpha\beta} \Omega^{AB} 
\epsilon_B^\beta = 0 \label{BPSspinor}
\ee
for a given symplectic form $\Omega_{AB}$ of $\mathds{C}^{2n}$, satisfying 
$\Omega_{AC} \Omega^{BC} = \delta_A^B$. Using (\ref{BPSspinor}) the `Dirac 
equation' (\ref{Dirac1}) can be further simplified to
\be
\scal{ a_A -  \Omega_{AB} a^B } | \C \rangle = 0 \label{Dirac} 
\ee
Let us designate by $\mathfrak{P}_{\frac{n}{\N}}$ the parabolic subgroup of $SO^*(2\N)$ that preserves the set of solutions of (\ref{BPSspinor}), and thus the set of supersymmetry charges leaving invariant a solution of Noether charge $|\C\rangle$ obeying (\ref{Dirac}). To derive  $\mathfrak{P}_{\frac{n}{\N}}$,  we note that the Lie algebra associated with $\mathfrak{P}_{\frac{n}{\N}}$ is generated by the eigen vectors of positive eigen value with respect to a $\gl_1$ generator $D_\frac{n}{\N} $ satisfying $D_\frac{n}{\N}  \epsilon^A_\alpha = \epsilon^A_\alpha$, which reads
\be 
D_\frac{n}{\N} \equiv \frac{1}{2} \scal{Ê\Omega_{AB} a^A a^B - 
\Omega^{AB} a_A a_B }  \in \so^*(2\N) 
\ee
in the harmonic oscillator basis (see \cite{nous} for details). Its
action on the spinor parameters satisfying (\ref{BPSspinor}) can be
read off from
\be [\, ÊD_\frac{n}{\N} \, , \, \epsilon_\alpha^A a_A + \varepsilon_{\alpha\beta} \epsilon_A^\beta a^A \, ] =   \, \epsilon_\alpha^A a_A + \varepsilon_{\alpha\beta} \epsilon_A^\beta a^A \ee
$D_\frac{n}{\N}$ defines a five-graded decomposition of 
$\so^*(2\N)$,
\begin{multline} \so^*(2\N) \cong  {\bf n(2n-1)}^\ord{-2} \oplus ( {\bf 2 n} \otimes {\bf 2 n_\bot})_\mathds{R}^\ord{-1} \\*   \oplus  \scal{\gl_1 \oplusÊ\su^*(2n) \oplus \so^*(2 n_\bot)}^\ord{0} \oplus ( {\bf 2 n} \otimes {\bf 2n_\bot})_\mathds{R}^\ord{1} \oplus  \overline{\bf n(2n-1)}^\ord{2} \end{multline}
with $n_\bot = \N - 2 n$. Indeed, using the identity,
\be [Ê\, D_\frac{n}{\N} \, ,\, a^A \pm  \Omega^{AB} a_B \, ] = \pm \scal{Êa^A \pm  \Omega^{AB} a_B } \ee
one concludes that the grade zero subalgebra is generated by
\bea \frac{1}{2}Ê{\Lambda_A}^B  [ \, a^A +  \Omega^{AC} a_C \, , \, a_B  +   \Omega_{BD} a^D \, ]Ê \hspace{3mm}  &\in& \hspace{3mm} \su^*(2n)^\ord{0} \CR
\frac{1}{2}Ê {\Lambda_{\bar A}}^{\bar B}  [ a^{\bar A}  , a_{\bar B}] + \frac{1}{2} \Lambda_{\bar A\bar B} a^{\bar A} a^{\bar B} - \frac{1}{2} \Lambda^{\bar A\bar B} a_{\bar A} a_{\bar B}  \hspace{3mm}  &\in& \hspace{3mm} \so^*(2n_\bot)^\ord{0}
 \eea
where $\bar A$ is a $U(n_\bot)$ index running from $2n+1$ to $\N$.
The nilpotent generators in $\so^*(2\N)$ of grade $\pm 1$ and
grade $\pm 2$ are, respectively,
\begin{multline} \Lambda_{A\bar B} \scal{Êa^A \pm  \Omega^{AC} a_C } a^{\bar B} - \Lambda^{A\bar B} \scal{Êa_A \mp  \Omega_{AC} a^C } a_{\bar B} \hspace{3mm}  \in \hspace{3mm}( {\bf 2 n} \otimes {\bf 2n_\bot})_\mathds{R}^\ord{\pm1} \\*
 \frac{1}{2} \Lambda_{AB}Ê\scal{Êa^A \pm  \Omega^{AC} a_C } \scal{Êa^B \pm  \Omega^{BD} a_D } - \frac{1}{2} \Lambda^{AB}Ê\scal{Êa_A \mp  \Omega_{AC} a^C } \scal{Êa_B \mp  \Omega_{BD} a^D } \hspace{30mm}  \\*
\in\hspace{3mm}  {\bf n(2n-1)}^\ord{\pm2}
\end{multline}
The covariance of the spinor parameter satisfying (\ref{BPSspinor}) under $\su^*(2n)$ follows from the commutation relations
\be
\big\{ a^A + \Omega^{AC} a_C \, , \, a^B + \Omega^{BD} a_D \big\} = 0 \hspace{10mm}  \big\{ a^A + \Omega^{AC} a_C \, , \, a_B + \Omega_{BD} a^D \big\} = 2 \delta^A_B \ee
The parabolic subgroup $\mathfrak{P}_\frac{n}{\N} \subset SO^*(2\N)$ preserving the set of spinor parameters  satisfying (\ref{BPSspinor}) is thus defined by the Lie algebra
\be \mathfrak{p}_\frac{n}{\N} \cong 
\scal{Ê \gl_1 \oplus \su^*(2n) \oplus \so^*(2 n_\bot)}^\ord{0} \oplus ( {\bf 2 n} \otimes {\bf 2n_\bot})_\mathds{R}^\ord{1} \oplus  \overline{\bf n(2n-1)}^\ord{2} \subset \so^*(2\N) 
 \ee
This group preserves the set of spinor parameters 
(\ref{BPSspinor}) in the sense that they are only rotated by the
subgroup $SU^*(2n)$ and rescaled by $GL(1,\mathds{R})$, while the remaining generators of $\mathfrak{P}_{\frac{n}{\N}}$
act trivially on them. In conclusion, acting on an \ft n\N \, BPS spherically 
symmetric solution with an element 
of $\mathfrak{P}_\frac{n}{\N} \times \H^*_0 $, one obtains 
another  \ft n\N \, BPS spherically symmetric solution which preserves 
the same supersymmetry charges.  

For asymptotically flat solutions, 
\ie asymptotically Taub--NUT space-times, the $\G$ Noether charge $\C$ 
is valued in $\g \ominus \h^*$. In supergravity, this is an irreducible 
representation of $Spin^*(2\N) \times \H^*_0$, that is, the product of the 
Weyl spinor representation ${\cal S}_+$ of $Spin^*(2\N)$ and an irreducible 
representation $\R$ of $\H^*_0$, possibly with an extra reality condition.

For simplicity let us first restrict ourselves to \ft 1\N \, BPS solutions. 
The theories admitting BPS solutions of higher BPS degree are quite 
restricted anyway and further relevant examples will be discussed 
in \cite{moi}. In this case the generator $D_\frac{1}{\N}$ 
defines the five graded decomposition (which is an example of the 
decomposition (\ref{Hfive}))
\be 
\so^*(2\N) \cong {\bf 1}^\ord{-2} \oplus ( {\bf 2 } \otimes 
{\bf 2 n_\bot})_\mathds{R}^\ord{-1} \oplus \scal{ \gl_1 \oplusÊ\su(2) 
\oplus \so^*(2 n_\bot)}^\ord{0} \oplus ( {\bf 2 } \otimes 
{\bf 2n_\bot})_\mathds{R}^\ord{1} \oplus  {\bf 1}^\ord{2} 
\ee
and the Weyl spinor representation of $Spin^*(2\N)$ decomposes accordingly as 
\be 
{\cal S}_+ \cong S_+^\ord{-1} \oplus ( {\bf 2} \otimes S_-)^\ord{0} 
\oplus S_+^\ord{1} 
\ee
where $S_\pm$ are the Weyl spinor representations of $SO^*(2 n_\bot)$.
Using the fact that the \ft 1 \N-BPS Killing spinors lie in the grade-one 
component of the vector representation of $SO^*(2\N)$,
\be 
{\bf 2\N}  \cong {\bf 2}^\ord{-1} \oplus {\bf 2n_\bot}^\ord{0} \oplus 
{\bf 2}^\ord{1} 
\ee
one sees that the solutions of the Dirac equations are simply obtained
by taking the charge matrix $\C$ to lie in the grade-one component of 
$\g\ominus \h^*$,
\be 
\C \in ( S_+ \otimes \R )^\ord{1} \subset {\cal S}_+ \otimes \R 
\ee 
It follows from the graded decomposition of $\g$ with respect to
$D_\frac{1}{\N}$ that for any charge matrices $\C_\n \in 
( S_+ \otimes \R )^\ord{1}$,
\be 
[ \C_\n , \C_\m ] \in {\bf 1}^\ord{2} \subset \so^*(2\N) \quad\Rightarrow\qquad
[ \C_\n , [ \C_\m , \C_\p ]] = 0 \label{commut} \ee 
In this way one obtains that such solution is of the type discussed in 
the preceding section, with ${\bf H} = 2 D_\frac{1}{\N}$. 

For $n\ge1$, the general solution of (\ref{Dirac})
is obtained by use of the identity
\be\label{Dirac2}
(a_A - \Omega_{AB} a^B) \exp \left( \frac12 \Omega_{CD} \,  a^C a^D \right)
  |0\rangle = 0
\ee
Consequently, if the charge $\C$ is represented by the state
\be 
|\C \rangle =  \biggl(\,  \Scal{Ê\w+ 
Z_{ij}Êa^i a ^j + \cdots } | 0 \rangle \,, \,  \Scal{Êz^\mathcal{A} + \Sigma^\mathcal{A}_{ij} a^i a^j + \cdots } | 0 \rangle \, \biggr)
\ee
the general \ft n\N-BPS solution reads
\be  
| \C \rangle =  \biggl(\,   e^{\frac{1}{2} \Omega_{AB} a^A a^B}   \Scal{Ê\w+ 
 Z_{\bar A \bar B} a^{\bar A} 
a^{\bar B}  } | 0 \rangle \,, \,  e^{\frac{1}{2} \Omega_{AB} a^A a^B} z^\mathcal{A} \,  | 0 \rangle \, \biggr)
\label{DiracSol} \ee
where $\bar A$ runs from $2n+1$ to $\N$, and $\mathcal{A}$ labels the
matter multiplets if $\N\leq 4$. An additional reality condition 
(Majorana--Weyl condition) may have to be imposed on the state 
if $\mathfrak{g} \ominus \mathfrak{h}^*$ admits one. It ensures that the scalar charges are completely determined in function of the others for any solution of the Dirac equation.\footnote{For instance for $\N= 2,\, 3$ the scalar charges are manifestly determined in (\ref{DiracSol}). For $\N=4$ the complex self-duality condition on the vector multiplets ensures that this is still the case, and similarly for $\N=8$ through the complex-selfduality of the multiplet.}  
In this case, the generator $\bf{H}$ is identified as
\be\label{H}
{\bf{H}}= \frac{2}{n} D_\frac{n}{\N}\, .
\ee
and $\bf{E}$ with the charge matrix $\C$ in such a way that the commutator
$[\bf{H},\bf{E}]$ corresponds to the action of the operator (\ref{H}) on 
the state $|\C\rangle$.\footnote{That $|\C\rangle$ is always
  null for BPS states obeying (\ref{Dirac}) then simply follows 
  from $\trace {\bf E}^2 = 0 $.} Because the component of grade two 
(with respect to $D_\frac{n}{\N}$) of $\so^*(2\N)$ leaves the 
{\ft n\N}-BPS Killing spinors invariant, the Killing equation trivially 
reduces to $d \epsilon^A_\alpha = 0$, and it follows from (\ref{Dirac}) 
that the corresponding supersymmetry variation of the dilatino 
fields vanishes. 

Having constructed the appropriate charge matrices $\C$ it is now 
straightforward to construct multi-black hole solutions of the 
four-dimensional theory following the procedure described in section~2. 
If the different matrices $\C_\m$ satisfy (\ref{Dirac1}) with respect
to the {\em same} supersymmetry parameters, the resulting multi-black
hole solution will also be supersymmetric. If the `constituent' black
holes are supersymmetric with respect to different supersymmetries, on
the other hand, the multi-black hole configuration will no longer be 
supersymmetric, and in general then exhibit naked singularities unless the `constituent' BPS black holes have a vanishing horizon area. Examples
of such solutions which are regular 
are given in \cite{moi}. Although supersymmetry is not required for 
the existence of Papapetrou--Majumdar type solutions, it greatly 
simplifies their explicit derivation. This is because the algebraic 
equations determining representatives of nilpotent orbits are not trivial 
to solve in general, whereas the linear Dirac equation (\ref{Dirac}) is. 
Nevertheless we will see that the supersymmetric graded decompositions 
can be used to derive the explicit form of the nilpotent 
charges in the non-supersymmetric case as well.

\section{Non-BPS solutions in five-dimensional theories}
The second class of solutions appear in the theories that can be obtained by toroidal compactification of 
(super)gravity theories in five dimensions to
four dimensions. Inspection of the list of theories displayed in \cite{Maison} shows that this is the case whenever $\H^*$ is semi-simple. The scalar fields of the five-dimensional theory then 
parametrise a symmetric space $\G_5 / \H_5$, and the isometry group 
$\G_5$ acts on the abelian vector fields in a faithful representation 
$\mathfrak{l}_5$. One then has the following three-graded decomposition 
of $\g_4$ and the associated four-graded decomposition of $\mathfrak{l}_4$,
\be 
\g_4 \cong \mathfrak{l}_5^\ord{-2}Ê\oplus \left(\gl_1 \oplus \g_5^\ord{0} 
\right) \oplus \mathfrak{l}_5^\ord{2} \hspace{10mm} \mathfrak{l}_4 \cong 
{\bf 1}^\ord{-3}Ê\oplus \mathfrak{l}_5^\ord{-1} \oplus  
\mathfrak{l}_5^\ord{1} \oplus  {\bf 1}^\ord{3} 
\ee
The maximal compact subalgeba $\h_4 \cong \h_5 \oplus \mathfrak{l}_5$ admits 
a non-compact real form $\h_4^* \cong   \h_5 \oplus  i \mathfrak{l}_5$ which 
appears in the pseudo-Riemmanian coset space $\G_4 / \H^*_4$ of the Euclidean 
four-dimensional theory obtained by dimensional reduction along the 
time-coordinate. Using the graded decomposition one obtains the following
result: because $\mathfrak{l}_4$ defines a complex irreducible representation of $\h_4 \subset \g_4$,
it decomposes into two conjugate real irreducible representations, 
$\mathfrak{l}^*_4$ and $\bar {\mathfrak{l}}^*_4$, of $\h_4^* \subset \g_4$. 
This property can be understood as follows. Let us define the four-dimensional
field strength $\mathcal{F}$ in the ${\bf 1} \oplus \mathfrak{l}_5$ 
representation, as well as  
$\mathcal{G}\equiv \delta{\mathcal{L}}/\delta{\mathcal{F}}$ (which in 
general depends on the scalar fields) such that the equations of motion of 
the abelian vector fields are $d \star \mathcal{G} = 0$. 
The $\H_4$-invariant form of the action on a four-dimensional 
pseudo--Riemannian space-time involves the complex combination 
$\star \mathcal{G} + i {\mathcal{F}}$ which transforms in the complex 
representation $\mathfrak{l}_4$ of $\H_4$. This representation is complex 
because $\star^2  = - 1$ for a pseudo-Riemanian metric, 
whereas the  $\H^*_4$-invariant form of the action on a four-dimensional 
Riemannian space involves the two independent {\em real} combinations 
$\star \mathcal{G} \pm\mathcal{F}$ which  transform in the real 
representation $\mathfrak{l}^*_4$ and its dual $\bar {\mathfrak{l}}^*_4$, 
respectively. As an example consider $\N=8$ supergravity, where 
$\star\mathcal{G} + i \mathcal{F}$ belongs to the complex  
representation $\bf{28}$ of $SU(8)$,
whereas in the Euclidean signature theory the corresponding combinations
$\star\mathcal{G} \pm \mathcal{F}$ belong to conjugate  {\em real} 
 representations $\bf{28}$ and $\overline{\bf{28}}$ of $SU^*(8)$.

It follows that when the four-dimensional theory descends from a 
five-dimensional one, the Lie algebra  $\h^* \cong \sl_2 \oplus 
\h_5 \oplus {\bf 3} \otimes \mathfrak{l}_5$ admits the following 
three-graded decomposition
\be 
\h^* \cong \mathfrak{l}_4^{*\ord{-2}} \oplus 
\left(\gl_1 \oplus \h_4^*\right)^\ord{0} \oplus  
\bar{\mathfrak{l}}_4^{*\ord{2}} 
\ee
which gives an example of decomposition (\ref{Hfive}), with $\h^\ord{4} 
\cong \emptyset$ in this case. Indeed one obtains the five-graded 
decomposition of $\g \ominus \h^*$,
\be 
\g \ominus \h^* \cong {\bf 1}^\ord{-4} \oplus  \bar{\mathfrak{l}}_4^{*\ord{-2}}
 \oplus ( \g_4 \ominus \h_4^*)^\ord{0} \oplus  \mathfrak{l}_4^{*\ord{2}} 
\oplus {\bf 1}^\ord{4} 
\ee
such that $ \mathfrak{l}_4^{*\ord{2}} \oplus {\bf 1}^\ord{4} $ defines an 
abelian nilpotent subalgebra of $\g$. One can thus define solutions of 
Papapetrou--Majumdar depending on $\dim[ \mathfrak{l}_5] + 2$ independent 
harmonic functions. As we will see, the most general asymptotically Minkowski 
solutions of this type are defined in function of charges lying in 
$ \mathfrak{l}_4^{*\ord{2}} $ and depend on $\dim[ \mathfrak{l}_5] + 1$ 
independent harmonic functions, and the extra-harmonic function corresponds 
to considering these extremal black holes in the background of pure 
NUT extremal black holes with vanishing horizon area.

\section{Maximal supergravity}
We now wish to illustrate the general discussion of the last sections 
through the example of the maximally supersymmetric supergravity. This 
theory being also the toroidal dimensional reduction of the maximally 
supersymmetric supergravity theory in five dimensions, it carries the 
two kinds of Papapetrou--Majumdar solutions discussed in the two preceding 
sections. They define the most general solutions of Papapetrou--Majumdar 
of the theory, which correspond to the only two $E_{8(8)}$ nilpotent 
orbits associated to the graded decomposition (\ref{fiveDbl}), 
respectively  \cite{nous,E8strat}. 

Let us first recall the form of the $E_{8(8)}$ Noether charge in the
harmonic oscillator basis \cite{nous}, with the mass and NUT charges 
$\w = m + i n$, the electromagnetic central charges $Z_{ij}$ and the 
scalar charges $\Sigma_{ijkl}$. The latter are determined as functions 
of $\w$ and $Z_{ij}$  through the characteristic equation (\ref{quintic}). 
The state $|\C\rangle\in\mathfrak{e}_{8(8)}\ominus\mathfrak{so}^*(16)$
is explicitly given by
\begin{multline}\label{128}  
| \C \rangle = \Scal{Ê\w + Z_{ij} a^i a^j + \Sigma_{ijkl} a^i a^j a^k a^l + \frac{1}{6!} \varepsilon_{ijklmnpq} Z^{pq} \, a^i \cdots  a^n +  \frac{1}{8!} \varepsilon_{ijklmnpq} \bar \w  \, a^i \cdots  a^q } | 0 \rangle \\*
= ( 1 + \invo ) \Scal{ÊÊ\w + Z_{ij} a^i a^j + 
\frac{1}{2}Ê\Sigma_{ijkl} a^i a^j a^k a^l } | 0 \rangle 
\end{multline} 
where $\invo$ is the anti-involution defining the 128-dimensional 
chiral Majorana--Weyl representations of $Spin^*(16)$ \cite{nous}. In the following subsections we 
will exhibit the form of $|\C\rangle$ explicitly for various kinds 
of extremal black holes.

\subsection{\ft18 BPS solutions}
For a generic \ft18 BPS solution preserving the 
supersymmetry generators associated to the Killing spinor parameters
\be 
\epsilon_\alpha^1 +  \varepsilon_{\alpha\beta} \epsilon^\beta_2 = 0 
\hspace{10mm} \epsilon_\alpha^{\bar A} = 0 
\ee
with $\bar A$ running from $3$ to $8$, the Noether charge $|\C\rangle$ must satisfy
\be\label{1/8BPS} 
\scal{Ê a^1 + a_2 } \, |\C\rangle =  \scal{Ê a^2 - a_1 } \, |\C\rangle = 0 
\ee 
The general solution can be written as
\be 
\C = ( 1 + \invo ) \Scal{Ê1 + a^1 a^2 } 
\scal{ \w + Z_{\bar A \bar B} a^{\bar A} a^{\bar B} } | 0 \rangle \label{eighthBPS}
\ee
This spinor evidently obeys the Majorana--Weyl condition 
$\invo|\C\rangle = |\C\rangle$, and to see that it still satisfies  
(\ref{1/8BPS}) in accordance with (\ref{Dirac2}) one uses the identity
\be\label{invo12}
( 1 + \invo ) \Scal{Ê1 + a^1 a^2 } 
\scal{ \w + Z_{\bar A \bar B} a^{\bar A} a^{\bar B} } | 0 \rangle 
=  \scal{Ê1 +  a^1 a^2 } ( 1 + \tilde\invo )  
\Scal{ \w + Z_{\bar A \bar B} a^{\bar A} a^{\bar B} } | 0 \rangle
\ee
where $\tilde\invo$ is the anti-involution for $\so^*(12)$.
The grading generator $D_\frac{1}{8}\in\mathfrak{so}^*(16)$ is 
\be D_\frac{1}{8} \equiv  a^1 a^2 - a_1 a_2 \ee
and obeys
\be D_\frac{1}{8} \, | \C \rangle = | \C \rangle  \ee
for any value of $Z_{\bar A \bar B}$. It gives rise to the following 
five-graded decomposition of $\so^*(16)$
\be 
\so^*(16) \cong {\bf 1}^\ord{-2} \oplus \scal{ {\bf 2} \otimes {\bf 12} }^\ord{-1}_\mathds{R} \oplus  \scal{\gl_1 \oplus Ê\su(2) \oplus \so^*(12) }^\ord{0} \oplus  \scal{ {\bf 2} \otimes {\bf 12} }^\ord{1}_\mathds{R} \oplus  {\bf 1}^\ord{2} 
\ee
The associated charge $\C\in\e_{8(8)} \ominus \so^*(16)$ 
is left invariant by $\su(2)$ and transforms as a Majorana--Weyl spinor 
${\bf{32}}_+$ of $\so^*(12)$, with the decomposition
\be 
\e_{8(8)} \ominus \so^*(16) \cong
{\bf 32}_+^\ord{-1} \oplus ({\bf 2} 
\otimes { \bf 32}_-)_\mathds{R}^\ord{0}Ê\oplus   {\bf 32}_+^\ord{1} 
\ee 
To see this explicitly, we invoke (\ref{invo12}) once more. 
The action of the grade zero component $\mathds{R}_+^* 
\times SU(2) \times Spin^*(12) \subset Spin^*(16)$, preserves the given 
supersymmetry charges by construction. In a similar way as for the action 
of $Spin^*(16)$ on the Kerr solutions, the moduli space of regular black 
holes preserving these four supersymmetry charges is the closure of 
the $\mathds{R}_+^* \times Spin^*(12)$ orbit of a given regular solution 
that can be chosen to correspond to the charge matrix 
$ \C = \scal{Ê1 +  a^1 a^2 } ( 1 + \tilde\invo ) | 0 \rangle $. This orbit is
\footnote{See \cite{nous} for our notations (which are not standard): 
  by `$\Ic$' we designate the parabolic extension of $SU(2)\times SU(6)$ 
  by the grade-one translation generators and the grade-two central charge
  \be \Ic\scal{ÊSU(2) \times SU(6)} \equiv \scal{ÊSU(2) \times SU(6)} 
  \ltimes \scal{ ( {\bf 2} \otimes {\bf 6})^\ord{1} 
  \oplus {\bf 1}^\ord{2}} \nn\ee
  Note also that the $SU(2)$ factors in the quotient 
  defining $\M_1^1$ in (\ref{orbit18}) `cancel out'.
  } 
\be 
\M_1^1\cong\frac{\mathds{R}_+^* \times Spin^*(12)}{SU(6)} \subset 
\M_1\cong\frac{ Spin^*(16)}{ \Ic ( SU(2) \times SU(6)) } \label{orbit18} 
\ee
This embedding is a refinement of the stratification (\ref{Strata}), 
with the following notation: by $\M_n^m$ for $m\leq n$ we 
generally designate the space of \ft n\N -BPS solutions which are such 
that $m$ out of $n$ supersymmetry charges are shared between all of 
its elements (which implies $\M_n^m\subset \M_n^{m'}$ for $m>m'$).
In other words, while $\M_1$ contains {\em all} \ft 18-BPS solutions,
irrespective of which supersymmetry charges are left invariant, $\M_1^1$
consists of all solutions that are \ft18 supersymmetric with respect to 
the {\em same} supersymmetry charge. Consequently, superposing solutions 
with charges $\C_\m$ from the orbit $\M_1^1$ according to the basic ansatz 
(\ref{Ansatz}) will result in a multi-black hole solution that still 
preserves \ft 18-supersymmetry. By contrast, taking different $\C_\m$ from 
the larger orbit $\M_1$ in general will lead to a solution that is no 
longer supersymmetric because the \ft18 supersymmetries of its black
hole `constituents' do not match.

Positivity of the energy entails that $|\w|$ is strictly greater than 
any of the eigenvalues of $Z_{\bar A \bar B}$, and that the quartic 
$E_{7(7)}$ invariant associated to it is strictly positive. Note that 
to avoid naked singularities, one must require moreover $\w \in \mathds{C} 
\setminus \mathds{R}_-$, and $\w\in \mathds{R}_+^*$ for asymptotically 
Minkowski solutions. The boundary of the orbit (\ref{orbit18}) decomposes 
into orbits corresponding to black holes with vanishing horizon 
area\footnote{where $ \Ic ( SU(2) \times Sp(3) \ltimes {\bf 14}_2) $ 
 is the contracted form of $\Ic( SU(2) \times SU(6))$ :
\be \scal{Ê SU(2) \times Sp(3) } \ltimes 
\scal{ Ê {\scal{Ê ({\bf 2}Ê\otimes {\bf 6} )_+ \oplus  {\bf 14}_2 }^\ord{1}  \oplus ({\bf 2}Ê\otimes {\bf 6} )_+^\ord{2} \oplus   {\bf 1}^\ord{3}} } \nn\ee}
\be\begin{array}{ccccccc}
\M_{1^\circ}^1&\cong& { \frac{Spin^*(12)}{Sp(3) \ltimes {\bf 14}_2}} &\subset& 
\M_{1^\circ}&\cong&\frac{ Spin^*(16)}{ \mathds{R}_+^*\ltimes \Ic 
( SU(2) \times Sp(3) \ltimes {\bf 14}_2) } \vspace{3mm} \\
\M_2^1&\cong& \frac{Spin^*(12)}{ (SU(2) \times Spin(1,6) ) \ltimes ( ( {\bf 2} \otimes {\bf 8})_\mathds{R} \oplus {\bf 1})} &\subset& \M_2&\cong&\frac{ Spin^*(16)}{ ( SU^*(4) \times Spin(1,6)) \ltimes ( ( {\bf 4} \otimes {\bf 8})_\mathds{R} \oplus {\bf 6}) } \vspace{3mm} \\
\M_4^1&\cong&\frac{ Spin^*(12)}{SU^*(6) \ltimes {\bf 15}} &\subset& \M_4&\cong&
\frac{ Spin^*(16)}{  SU^*(8)\ltimes {\bf 28}} 
\end{array}
\ee
with the notation from \cite{nous} (see especially eq.~(5.17) there)
for the embedding of the various `BPS strata' into the stratified space of 
regular extremal solutions of the characteristic equation (\ref{quintic}).
So in particular, and in accordance with the notation introduced above, 
the `BPS substrata' $\M_2^1$ and $\M_4^1$ correspond to \ft14 and 
\ft12 BPS black holes, all of which share the same \ft18 supersymmetry, 
respectively (while by $\M_{1^\circ}$ we denote the stratum of general
\ft18 BPS solutions of vanishing horizon area). In all
three cases the $E_{7(7)}$ invariant $\lozenge(\w^{-\frac{1}{2}} Z)$ 
associated to $\C$ vanishes, and zero, one or three of the eigenvalues 
of $Z_{\bar A\bar B}$ are equal to $\w$ in modulus, respectively. These 
orbits correspond to the various Majorana--Weyl spinors of null type 
which turn out to be slightly more complicated than in the case of the 
null vectors of $SO(p,q)$. For example, the $Spin^*(12)$ Majorana--Weyl
spinors associated to \ft12 BPS solutions are pure in the sense of Cartan. 

By virtue of the grading, any linear combination of these charges 
satisfies $\C^5 = 0$, and moreover, the commutator of any two of them gives 
\be 
[  \C_\n , \C_\m ] = \scal{Ê\w_\n \bar \w_\m -  \bar \w_\n  \w_\m - 
2 Z_{\n\, \bar A \bar B} Z_\m^{\bar A \bar B} + 
2 Z_\n^{\bar A \bar B} Z_{\m\, \bar A \bar B} }Ê( a^1 + a_2 ) 
( a^2 - a_1 ) \in {\bf 1}^\ord{2} 
\ee
The generator $ Ê( a^1 + a_2 ) ( a^2 - a_1 ) $ of $\so^*(16)$ leaves 
invariant the Killing spinor associated to the charges $\C_\n$, and 
the Killing spinor equation trivially reduces to
\be d \epsilon^A_\alpha = 0 \ee
We have thus derived the most general \ft18 BPS solution of maximal 
supergravity, which depends on $32$ independent harmonic functions 
associated to the independent components of the $Spin^*(12)$ Majorana--Weyl 
spinor.  Of course in general, some of the $\C_\n$ can preserve more 
supersymmetry, in such a way that each black hole can be either a generic 
\ft18 BPS black hole carrying one of the $32$ linearly independent charges of $\M_1^1$, or a \ft18 BPS black hole of vanishing horizon 
area carrying one of the $31$ linearly independent charges of $\M_{1^\circ}^1$, 
or a \ft14 BPS black hole carrying one of the $25$ linearly independent 
  charges of $\M^1_2$, or a \ft12 BPS black hole carrying one of the 
$16$ linearly independent charges of $\M^1_4$. All these solutions can 
be understood as \ft12 BPS multi-black holes of the magic $\N=2$ 
supergravity associated to the quaternions \cite{Kallosh}. Within the 
latter truncation, $D_\frac{1}{8}$  is a non compact generator of the $\sl_2$ 
component of  $\sl_2 \oplus \so^*(12) \subset \e_{7(-5)}$, which 
decomposes $\e_{7(-5)}$ as 
\be 
\mathfrak{e}_{7(-5)} \cong  {\bf 1}^\ord{-2} \oplus {\bf 32}_+^\ord{-1} 
\oplus \scal{Ê\mathfrak{gl}_1 \oplus \so^*(12)}^\ord{0} \oplus {\bf 32}_+^\ord{1} 
\oplus  {\bf 1}^\ord{2}  \label{SOXII} 
\ee
exhibiting the fact that these solutions can be understood inside 
the $\N=2$ truncated theory.

\subsection{\ft14 BPS solutions}
For solutions preserving \ft14 supersymmetry the charge
state is of the form (using (\ref{Dirac2}) once again) 
\be 
|\C \rangle = e^{\frac{1}{2} \Omega_{AB} a^A a^B} \, \Scal{Ê\w + Z_{\bar A \bar B} a^{\bar A} a^{\bar B} + \frac{1}{24} \varepsilon_{\bar A \bar B \bar C \bar D } \, \bar \w\, a^{\bar A} a^{\bar B} a^{\bar C} a^{\bar D} } | 0 \rangle  
\label{quarter} \ee
in an appropriate basis, where $\Omega_{AB}$ is a complex-selfdual symplectic 
form of $\mathds{C}^4$ and $Z_{\bar A\bar B}$ is complex-selfdual with 
respect with the complementary $U(4)$. The expression in parantheses is
thus a chiral spinor of $Spin^*(8)$; by triality and because of the 
isomorphism $Spin^*(8)\cong Spin(2,6)$ it can be equivalently viewed as 
a {\em vector} of $SO(2,6)$. The Killing spinors for (\ref{quarter}) 
then satisfy
\be 
\epsilon^A_\alpha + \varepsilon_{\alpha\beta} \Omega^{AB} 
\epsilon^\beta_B = 0 \hspace{20mm} \epsilon_\alpha^{\bar A} = 0 
\ee 
The associated grading operator of $\so^*(16)$ is 
\be D_\frac{1}{4} \equiv \frac{1}{2} \scal{Ê\Omega_{AB} a^A a^B - \Omega^{AB} a_A a_B} \ee
and we now have
\be D_\frac{1}{4} \,  |\C \rangle = 2\,  | \C \rangle \ee 
The corresponding five-graded decomposition of $\so^*(16)$ is 
\be \so^*(16) \cong {\bf 6}^\ord{-2} \oplus \scal{Ê{\bf 4} \otimes {\bf 8}}_\mathds{R}^\ord{-1} \oplus \scal{Ê  \gl_1 \oplus \su^*(4) \oplus \so^*(8) }^\ord{0} \oplus  \scal{Ê{\bf 4} \otimes {\bf 8}}_\mathds{R}^\ord{1} \oplus {\bf 6}^\ord{2} 
\ee
One computes that $|\C\rangle$ transforms as a Majorana--Weyl $Spin^*(8)$ 
spinor with respect with $SU^*(4) \times Spin^*(8)\cong Spin(5,1) \times Spin(2,6)$,
 \ie as a vector of $SO(2,6)$. Indeed
\be\label{grad14} 
\e_{8(8)} \ominus \so^*(16) \cong {\bf 8}_+^\ord{-2} \oplus ( {\bf 4} \otimes {\bf 8}_- )_\mathds{R}^\ord{-1} \oplus ({\bf 6} \otimes { \bf 8}_+ )^\ord{0}Ê\oplus ( {\bf 4} \otimes {\bf 8}_- )_\mathds{R}^\ord{1} \oplus   {\bf 8}_+^\ord{2}\ee 
where by ${\bf{8}}$ and ${\bf{8}}_\pm$ we denote the three inequivalent
fundamental representations of $Spin^*(8)$.
In accordance with the grading (\ref{grad14}) a charge matrix of the 
form (\ref{quarter}) transforms as ${\bf{8}}_+$ and defines 
a nilpotent abelian subalgebra $\mathds{R}^{2+6}$ of $\mathfrak{e}_{8(8)}$. 
The corresponding moduli space is
\be\label{M22}
\M_2^2\cong \frac{ \mathds{R}_+^* \times SO(2,6)}{SO(1,6)} \subset 
\M_2\cong \frac{Spin^*(16)}{ ( SU^*(4) \times Spin(1,6) )
\ltimes ( ({\bf 4} \times {\bf 8})_\mathds{R} \oplus {\bf 6})} 
\ee
The subgroup $SO(1,6)$ here is the isotropy group of a timelike vector
of $SO(2,6)$. In order for the corresponding multi-black holes solutions 
to be regular, one must thus require the charge matrices in (\ref{M22})
to correspond to non-space-like vectors. When the norm of the $SO(2,6)$ 
vector goes to zero, we reach the boundary of $\M_2^2$ corresponding to the
subspace of \ft12 BPS charge matrices belonging to
\be  
\M_4^2\cong \frac{  SO(2,6)}{ISO(1,5)} \subset 
\M_4\cong\frac{Spin^*(16)}{ SU^*(8) \ltimes {\bf 28} }
\ee
where we again employ the notation in (\ref{orbit18}).
So we conclude that the general \ft14 BPS solutions depend on eight 
independent harmonic functions associated to the linearly independent 
non-spacelike $SO(2,6)$ vectors. Note nevertheless that even if all the 
vectors are chosen to be null, the solution only preserves one quarter 
supersymmetry since, although each black hole is \ft12 BPS, there are 
only eight supercharges that are preserved by each one of them.  As the 
most general \ft18 BPS solutions were possibly understood within the 
magic $\N=2$ truncated theory associated to the quaternions, the general 
\ft14 BPS solutions can be understood within $\N=4$ supergravity coupled 
to six vector multiplets. In the latter truncation, $D_\frac{1}{4}$ is the 
generator of the $\so^*(8)$ R-symmetry group of the theory,
\be 
\so^*(8) \cong {\bf 6}^\ord{-2} \oplus \scal{ \gl_1 \oplus \su^*(4)}^\ord{0} \oplus  {\bf 6}^\ord{2} 
\ee
such that 
\be 
\so(8,8) \ominus \scal{ \so^*(8) \oplus \so^*(8) } \cong  {\bf 8}_+^\ord{-2} 
\oplus ( {\bf  6} \otimes {\bf 8}_+)^\ord{0} \oplus  {\bf 8}_+^\ord{2} 
\ee

\subsection{\ft12 BPS solutions}
For \ft12 BPS solutions, the set of $16$ supersymmetry charges is associated 
to the Killing spinors 
\be 
\epsilon_\alpha^i + \varepsilon_{\alpha\beta} \Omega^{ij} 
\epsilon^\beta_j  = 0 
\ee
where $\Omega_{ij}$ is a (generally complex) symplectic form of 
$\mathds{C}^8$ satisfying $\Omega_{ik} \Omega^{jk} = \delta_i^j$
(with our usual convention $\Omega^{ij}\equiv (\Omega_{ij})^*$).
In this case the four-dimensional R-symmetry group $SU(8)$ permits 
to rotate $\Omega_{ij}$ to a diagonal basis 
\be 
\Omega_{ij} \,  \hat{=} \, e^{i\omega} \, \left( \begin{array}{cc} 0 &\, 
\,  \mathds{1}  \\ - \mathds{1} &\, \, 0 \end{array}\right) 
\ee
but one cannot eliminate the overall phase $\omega$ in general. 
Eq.~(\ref{Dirac}) uniquely determines the charge matrix as
\be 
| \C \rangle = m e^{-2i\omega} \,  e^{\frac{1}{2} 
\Omega_{ij} a^i a^j} \, | 0 \rangle 
\ee
where the overall phase factor $e^{-2i\omega}$ ensures that $|\C\rangle$
is a Majorana spinor for any choice of $\omega$ via the identity
\be 
e^{-2i \omega} \Omega_{[ij}   \Omega_{kl]}  
= e^{2i \omega} \frac{1}{24} \varepsilon_{ijklmnpq} \Omega^{mn}\Omega^{pq} 
\ee
Asymptotically Minkowskian \ft12 BPS black holes thus correspond to 
$\omega=0$. The asymptotic central charges are then purely electric 
modulo an $SU(8)$ rotation.

The general \ft 12 BPS solution is determined by one single harmonic function 
and the given \ft12 BPS charge matrix. The relevant grading operator is 
\be\label{D1/2} 
D_\frac{1}{2} \equiv \frac{1}{2} \scal{Ê\Omega_{ij} a^i a^j 
- \Omega^{ij} a_i a_j} 
\ee
and satisfies $D_\frac{1}{2}\, | \C \rangle = 4  | \C \rangle$. It commutes 
with the generators of $\su^*(8)$ and its action on the nilpotent generators 
follows from
\be  [Ê\, D_\frac{1}{2} \, ,\, a^i \pm  \Omega^{ij} a_j \, ] = \pm Ê \scal{Êa^i \pm  \Omega^{ij} a_j } \ee
and thus gives rise to the decomposition of $\so^*(16)$
\be \so^*(16) \cong {\bf 28}^\ord{-2} \oplus 
\scal{Ê Ê\gl_1 \oplus \su^*(8) }^\ord{0} \oplus \overline{\bf 28}^\ord{2} 
\ee
such that 
\be
\e_{8(8)} \ominus \so^*(16) \cong {\bf 1}^\ord{-4}Ê\oplus 
\overline{\bf 28}^\ord{-2} \oplus {\bf 70}^\ord{0} \oplus  
{\bf 28}^\ord{2} \oplus {\bf 1}^\ord{4} \label{su8}
\ee
and $|\C\rangle\equiv |\C^\ord{4}\rangle\in {\bf{1}}^\ord{4}$. Let us
point out once again that the level $\pm 2$ representations of $SU^*(8)$
are both {\em real}, but dual to one another, and correspond to 
{\em independent} real combinations of the field strengths, as 
explained in section~5.

\subsection{Non-BPS solutions}
The results of the preceding section can be summarised by characterising 
the relevant vector space of $\g \ominus \h^*$, \ie $\g^\ord{2} \ominus 
\h^\ord{2} \cong {\bf 32}, \,  {\bf 8}$ and ${\bf 1}$, for the \ft18 BPS, 
the \ft14 BPS and the \ft12 BPS solutions, respectively. One can now
construct non-BPS multi-particle solutions in a similar way. Namely, the
charge matrices of non-BPS extremal solutions are associated to the same 
graded decomposition (\ref{su8}) of $\e_{8(8)}$  as the maximally 
supersymmetric ones. The only difference is that the associated 
charge matrices now belong to the grade one component of (\ref{su8}), 
which defines an abelian nilpotent subalgebra $\mathds{R}^{1+27}\subset 
\e_{8(8)}$. The relevant ($\omega$-dependent) generator ${\bf H}  
\in\so^*(16)$  is thus identical with $D_\frac{1}{2}$, cf.~(\ref{D1/2}). 

The simplest solution of this type is obtained from the state
\be 
| \C^\ord{2}Ê\rangle =  
( 1 + \invo )  \Scal{Ê1 + \frac{1}{4} \Omega_{ij} a^i a^j }Ê  i e^{-2i \omega}Ê  | 0 \rangle \ee
where the overall phase is determined such that the four-form component of
\be 
\scal{Ê {\bf H}  - 2 }Ê| \C^\ord{2}Ê\rangle \Big|_{4\!-\!{\rm form}} =  
\frac{1}{8}\,  (1+\invo) \, i e^{-2i \omega}\,  \Omega_{ij} \Omega_{kl}\,  
a^i a^j a^k a^l |0\rangle =0  
\ee 
vanishes due to the action of the projection operator $\frac{1}{2}(1+\invo)$. 
In order 
to obtain non-BPS asymptotically Minkowskian extremal black holes, one must 
therefore choose $\omega  = \frac{\pi}{4}$. The top grade component then becomes associated to a pure \ft12 BPS Taub--NUT 
black hole as would be obtained from the level four charge matrix
\be | \C^\ord{4}\rangle  = i n \, e^{ \frac{1}{2}Ê\Omega_{ij}  a^i a^j } 
\, | 0\rangle \label{NUT} 
\ee
The charge matrix of grade two depends on the associated mass 
$m$ and a rank two tensor $Q_{ij}$ satisfying\footnote{Which means that 
  $Q_{ij}$ lies in the ${\bf 27}$ representation of $Sp(4)\subset SU(8)$ 
  which leaves invariant $\Omega_{ij}$.}
\be 
Q_{ij} = \Omega_{ik} \Omega_{jl} Q^{kl} \hspace{20mm}Ê\Omega^{ij} 
Q_{ij} = 0 \label{real27} 
\ee
such that
\be | \C^\ord{2} \rangle  = ( 1 + \invo )  \Scal{Ê1 + \frac{1}{4} 
\Omega_{ij} a^i a^j }Ê\scal{Ê m + Q_{ij} a^i a^j } |0\rangle  \label{nonBPS}
\ee
In this way one obtains multi-particle solutions depending on $28$ 
harmonic functions. Although the symplectic form is not an invariant 
tensor of $SU^*(8)$, the reality condition
\be 
Z_{ij} = \Omega_{ik} \Omega_{jl} Z^{kl} 
\ee
is nonetheless preserved by $SU^*(8)$, and it defines the real ${\bf 28}$ 
representation of $SU^*(8)$. This is also the representation to which
belong the asymptotic central charges 
\be Z_{ij} = \frac{m}{4} \Omega_{ij} + Q_{ij} \ee
In contradistinction to the \ft12 BPS asymptotically Minkowski black holes, 
the non-BPS ones are therefore purely dyonic ($\omega = \frac{\pi}{4}$), 
up to an $SU(8)$ rotation.
 
$|\C^\ord{2}\rangle$ corresponds to a regular black holes as long as 
the eigenvalues of $Z_{ij}$ are less than or equal to $m$ in 
modulus. The $E_{7(7)}$ quartic invariant $\lozenge(Z)$ is then negative, 
\be 
\lozenge(Z) = - \frac{m}{16}^{\hspace{-0.3mm}4} + 
\frac{m}{4}^{\hspace{-0.3mm}2}  Q^{ij} Q_{ij} + 
\frac{m}{6} \Omega_{ij} Q^{jk} Q_{kl} Q^{li} - 
\frac{1}{2} \scal{ÊQ^{ij} Q_{ij} }^2 + 2 Q_{ij} Q^{jk} Q_{kl} Q^{li} \le 0
\ee
In order to spell out the these conditions explicitly, it is 
useful to find a basis in which both $\Omega_{ij}$ and $Q_{ij}$ are block 
diagonal (which always exists thanks to the reality condition (\ref{real27}))
\be \Omega_{ij} \hat{=} e^{\frac{i \pi}{4}} \, \left( \begin{array}{cc} 0 &\, \,  1  \\ - 1 &\, \, 0 \end{array}\right) \otimes \mathds{1} \hspace{8mm}ÊQ_{ij} \hat{=} \frac{e^{\frac{i \pi}{4}}}{2}Ê\, \, \left( \begin{array}{cc}0 & \, \, 1  \\- 1 & \, \, 0 \end{array}\right) \otimes  \left( \begin{array}{cccc} \, \, \rho_\un \, \, Ê& 0 & 0 & 0 \\0& \, \, \rho_\deux \, \,  & 0 & 0 \\ \, \, 0 & 0 &\, \,  \rho_\trois \, \, & 0 \\ 0 & 0 & 0 & { \scriptstyle - \rho_\un - \rho_\deux - \rho_\trois }  \end{array} \right) \ee
The three real parameters $\rho_{\mathpzc{i}}$ are constrained to lie in 
the tetrahedron defined by the conditions $\rho_{\mathpzc{i}} \le 
\frac{m}{2}Ê$ for ${\mathpzc{i}} = 1 ,\, 2 ,\, 3$ and $ - \frac{m}{2} 
\le \rho_\un + \rho_\deux + \rho_\trois$. In this basis, 
the $E_{7(7)}$ invariant reads 
\be \lozenge(Z) = - \left( \frac{m}{2} - \rho_\un \right) 
\left( \frac{m}{2} - \rho_\deux \right) \left(\frac{m}{2} - \rho_\trois \right) \left( Ê\frac{m}{2} + \rho_\un + \rho_\deux + \rho_\trois \right)
\ee
and one checks easily that it is negative inside the tetrahedron 
and that it vanishes on its faces. In the absence of NUT charge, 
one expects the horizon area of such black holes to be given by
\be A_\mathscr{H} =  4 \pi \sqrt{Ê - \lozenge(Z)}   \ee
Such multi-black hole solutions include non-BPS black holes corresponding 
to one of the $28$ linearly independent charge matrices in 
\be \frac{\mathds{R}_+^* \times SU^*(8)}{Sp(4)} \subset 
\frac{ Spin^*(16)}{Sp(4) \ltimes {\bf 27}} \ee
(for which the parameters $\rho_{\mathpzc{i}}$ lie inside the tetrahedron)
as well as \ft18 BPS black holes corresponding to one of the $27$ linearly 
independent charge matrices in  
\be \frac{SU^*(8)}{\scal{Ê SU(2) \times Sp(3)}\ltimes ({\bf 2} 
\otimes {\bf 6})_\mathds{R}} \subset \frac{ Spin^*(16)}
{\mathds{R}_+^*\ltimes \Ic ( SU(2) \times Sp(3) \ltimes {\bf 14}_2)} 
\ee
(for which $\rho_{\mathpzc{i}}$ lie on a face of the tetrahedron).
Finally, \ft14 BPS black holes correspond to one of the $ 22$ linearly 
independent charge matrix in 
\be
\frac{SU^*(8)}{ \scal{ÊSU^*(4)\times Sp(2)  } \ltimes ( {\bf 4} \otimes {\bf 4})_\mathds{R}} \subset \frac{ Spin^*(16)}{ \scal{Spin(5,1) \times Spin(1,6)} \ltimes ( ( {\bf 4} \otimes {\bf 8})_\mathds{R} \oplus {\bf 6}) } 
\ee
(with $\rho_{\mathpzc{i}}$ lying on an edge of the tetrahedron)
and \ft12 BPS black holes to one of the $13+1$ linearly independent 
charge matrices in 
\be
\frac{ SU^*(8)}{ \scal{ÊSU^*(6) \times Sp(1)  } \ltimes ( {\bf 6} \otimes {\bf 2 })_\mathds{R}}   \cup \mathds{R} \subset \frac{ Spin^*(16)}{  SU^*(8)\ltimes {\bf 28}} 
\ee
(corresponding to $\rho_{\mathpzc{i}}$ lying on a vertex of the tetrahedron).
This set includes the charge matrix (\ref{NUT}) lying in the component 
of grade four ${\bf 1}^\ord{4}$.

Note that these solutions are completely different from the BPS ones. To see 
this, we note that the decomposition of $\e_{8(8)}$ associated to a \ft12 BPS 
charge matrix within the graded decomposition associated to either the 
\ft18 BPS solutions or the non-BPS extremal solutions lead to equivalent 
decomposition in terms of representations of $SU(2) \times SU^*(6)$. The 
charge matrix in the neighbourhood of this charge which define \ft12 BPS black holes which do not interact with the \ft12 BPS black hole associated to the latter, can be obtained by acting with either $\so^*(12)$ or $\su^*(8)$, for \ft18 BPS and non-BPS solutions, respectively. The action of $\so^*(12)$ generates new components in the ${\bf 15}$ of $SU^*(6)$ whereas the action of $\su^*(8)$ generates new components in the $({\bf 2} \otimes {\bf 6})_\mathds{R}$ of $SU(2) \times SU^*(6)$, which shows that they are inequivalent. 

\vspace{0.5cm}

\noindent
{\bf Acknowledgments}: We are grateful to Boris Pioline and Kelly Stelle
for discussions and comments.




\begin{thebibliography}{99}


\bibitem{Papapetrou}
A.~ Papapetrou,
``A static solution of the equations of the gravitational field for an arbitrary charge distribution,''
Proc.\ Roy.\ Irish Acad.\ A {\bf 51}, 191 (1945).

\bibitem{Majumdar}
S.~D.~Majumdar,
``A class of exact solutions of Einstein's field equations,''
Phys.\ Rec.\ {\bf 72}, 390 (1945).


\bibitem{Denef}
  B.~Bates and F.~Denef,
  ``Exact solutions for supersymmetric stationary black hole composites,''
  \eprint{hep-th/0304094}.


\bibitem{attractors}
 S.~Ferrara, R.~Kallosh and A.~Strominger,
  ``$\N=2$ extremal black holes,''
  Phys.\ Rev.\  D {\bf 52}, 5412 (1995)
  \eprint{hep-th/9508072}.
  
 \bibitem{attractors1}
  S.~Ferrara and R.~Kallosh,
  ``Supersymmetry and attractors,''
  Phys.\ Rev.\  D {\bf 54}, 1514 (1996)
  \eprint{hep-th/9602136}.

\bibitem{Maison}
  P.~Breitenlohner, D.~Maison and G.~W.~Gibbons,
  ``Four-dimensional black holes from Kaluza--Klein theories,''
  Commun.\ Math.\ Phys.\  {\bf 120}, 295 (1988).

\bibitem{Maison1} 
P.~Breitenlohner and D.~Maison,
  ``On nonlinear sigma-models arising in \\ (super-)gravity,''
  Commun.\ Math.\ Phys.\  {\bf 209}, 785  (2000)
  \eprint{gr-qc/9806002}.


\bibitem{nous}
  G.~Bossard, H.~Nicolai and K.~S.~Stelle,
  ``Universal BPS structure of stationary supergravity solutions,''
 \eprintN{0902.4438}.

\bibitem{Clement}
  G.~Clement and D.~V.~Galtsov,
  ``Stationary BPS solutions to dilaton-axion gravity,''
  Phys.\ Rev.\  D {\bf 54}, 6136 (1996)
  \eprint{hep-th/9607043}.


\bibitem{Boris}
  M.~Gunaydin, A.~Neitzke, B.~Pioline and A.~Waldron,
  ``BPS black holes, quantum attractor flows and automorphic forms,''
  Phys.\ Rev.\  D {\bf 73}, 084019 (2006)
  \eprint{hep-th/0512296}.


\bibitem{Gaiotto}
  D.~Gaiotto, W.~W.~Li and M.~Padi,
  ``Non-supersymmetric attractor flow in symmetric spaces,''
  JHEP {\bf 0712}, 093  (2007) 
  \eprintN{0710.1638}.



\bibitem{CJ} 
  E.~Cremmer and B.~Julia, 
  ``The $SO(8)$ supergravity,''
  {Nucl.\ Phys.\ B} {\bf159}, 141 (1979).


\bibitem{moi}
G.~Bossard,
``The extremal black holes of $\N=4$ supergravity from $\so(8,2+n)$ 
nilpotent orbits'', to appear.

\bibitem{Ferrara}
  S.~Bellucci, S.~Ferrara, M.~Gunaydin and A.~Marrani,
  ``SAM lectures on extremal black holes in $d=4$ extended supergravity,''
  \eprintN{0905.3739}.

\bibitem{coadjoint}
D.~H.~Collingwood and W.~M.~McGovern,
{\it Nilpotent orbits in semisimple Lie algebra,}
Van Nostrand Reinhold mathematics series, New York  (1993).

\bibitem{DallAgatta}
G.~Lopes Cardoso, A.~Ceresole, G.~Dall'Agata, J.~M.~Oberreuter and J.~Perz,
  ``First-order flow equations for extremal black holes in very special
  geometry,''
  JHEP {\bf 0710}, 063 (2007)
  \eprintN{0706.3373}.


\bibitem{Hotta}
  K.~Hotta and T.~Kubota,
  ``Exact solutions and the attractor mechanism in non-BPS black holes,''
  Prog.\ Theor.\ Phys.\  {\bf 118}, 969 (2007)
   \eprintN{0707.4554}.

\bibitem{Gimon}
  E.~G.~Gimon, F.~Larsen and J.~Simon,
  ``Black holes in supergravity: the non-BPS branch,''
  JHEP {\bf 0801}, 040 (2008) 
  \eprintN{0710.4967}.

\bibitem{Breit}
P.~Breitenlohner and D.~Maison,
``Solitons in Kaluza--Klein theories''
{\it Solitons in general relativity}, H.~Morris and R.~Dodd eds.\  (1986).


\bibitem{Book} H.~Stephani, D.~Kramer, M.A.~MacCallum, C.~Hoenselars and 
   E.~Herlt, {\it Exact solutions of Einstein's field equations},
   Cambridge University Press (2003)
   
\bibitem{magic}
 M.~G\"unaydin, G.~Sierra and P.~K.~Townsend,
  ``Exceptional supergravity theories and the magic square,''
  Phys.\ Lett.\  B {\bf 133}, 72 (1983).

\bibitem{Kallosh}
  S.~Ferrara, E.~G.~Gimon and R.~Kallosh,
  ``Magic supergravities, $\N = 8$ and black hole composites,''
  Phys.\ Rev.\  D {\bf 74}, 125018 (2006)
  \eprint{hep-th/0606211}.

\bibitem{E8strat}
D.~\v{Z}.~\DJo,
``The closure diagram for nilpotent orbits of the split real form of $E_8$,''
CEJM {\bf 4}, 573 (2003).

%
\end{thebibliography}
\end{document}